\documentclass[lettersize,journal]{IEEEtran}
\usepackage{amsmath,amsfonts}
\usepackage{algorithmic}
\usepackage{algorithm}
\usepackage{array}
\usepackage[caption=false,font=normalsize,labelfont=sf,textfont=sf]{subfig}
\usepackage{textcomp}
\usepackage{stfloats}
\usepackage{url}
\usepackage{verbatim}
\usepackage{graphicx}
\usepackage{cite}
\usepackage{enumerate}
\usepackage{amsmath,amssymb,amsfonts}
\usepackage{algorithmic}
\usepackage{graphicx}
\usepackage{textcomp}
\usepackage{xcolor}
\usepackage{hyperref}
\usepackage{graphicx}
\usepackage{float}
\usepackage{subfig}
\usepackage{amsthm}
\usepackage{mathrsfs}

\usepackage{pdfpages}
\usepackage{colortbl}
\definecolor{mygray}{gray}{0.85}
\usepackage{array}
\usepackage{caption}
\usepackage{setspace}
\usepackage{diagbox}
\usepackage[figuresright]{rotating}
\definecolor{mygray}{rgb}{0.95,0.99,0.95}
\definecolor{mypink}{rgb}{.99,.93,.85}
\definecolor{mycyan}{cmyk}{.2,0.04,0,0}
\usepackage{xcolor}  
\usepackage{colortbl,booktabs}
\usepackage{multirow}
\usepackage{threeparttable}

{}
{}
{}

\newcolumntype{I}{!{\vrule width 1.25pt}}
\newlength\savedwidth

\newlength\savewidth
\newcommand\shline{\noalign{\global\savewidth\arrayrulewidth
		\global\arrayrulewidth 1.25pt}%
	\hline
	\noalign{\global\arrayrulewidth\savewidth}}
\begin{document}

\title{AI-Empowered Multiple Access for 6G: A Survey of Spectrum Sensing, Protocol Designs, and Optimizations}

\author{
\IEEEauthorblockN{Xuelin Cao, Bo Yang, Kaining Wang, Xinghua Li,
  Zhiwen Yu,~\IEEEmembership{Senior member,~IEEE},\\
  Chau Yuen,~\IEEEmembership{Fellow,~IEEE},
  Yan Zhang,~\IEEEmembership{Fellow,~IEEE},
  Zhu Han, \IEEEmembership{Fellow,~IEEE}
  }
}
\markboth{Journal of \LaTeX\ Class Files,~Vol.~14, No.~8, August~2021}%
{Shell \MakeLowercase{\textit{et al.}}: A Sample Article Using IEEEtran.cls for IEEE Journals}


\maketitle

\begin{abstract}
With the rapidly increasing number of bandwidth-intensive terminals capable of intelligent computing and communication, such as smart devices equipped with shallow neural network models, the complexity of multiple access for these intelligent terminals is increasing due to the dynamic network environment and ubiquitous connectivity in 6G systems. Traditional multiple access (MA) design and optimization methods are gradually losing ground to artificial intelligence (AI) techniques that have proven their superiority in handling complexity. AI-empowered MA and its optimization strategies aimed at achieving high Quality-of-Service (QoS) are attracting more attention, especially in the area of latency-sensitive applications in 6G systems. In this work, we aim to: 1) present the development and comparative evaluation of AI-enabled MA; 2) provide a timely survey focusing on spectrum sensing, protocol design, and optimization for AI-empowered MA; and 3) explore the potential use cases of AI-empowered MA in the typical application scenarios within 6G systems. Specifically, we first present a unified framework of AI-empowered MA for 6G systems by incorporating various promising machine learning techniques in spectrum sensing, resource allocation, MA protocol design, and optimization. We then introduce AI-empowered MA spectrum sensing related to spectrum sharing and spectrum interference management. Next, we discuss the AI-empowered MA protocol designs and implementation methods by reviewing and comparing the state-of-the-art, and we further explore the optimization algorithms related to dynamic resource management, parameter adjustment, and access scheme switching. Finally, we discuss the current challenges, point out open issues, and outline potential future research directions in this field.
\end{abstract}


\begin{table*}[!htb]
\centering
\caption{List of abbreviations}
%\small
\begin{tabular}{|l l||l l|}
\shline
ABS& Almost Blank Sub-Frame & MEC & Multi-Access Edge Computing\\
ACL&Actor-Critic Learning & MABL & Multi-Armed Bandit Learning\\
AIFS&Arbitration Inter-Frame Space & MA & Multiple Access\\
AI& Artificial Intelligence& MIMO&Multiple-Input Multiple-Output\\
ANN& Attention-Based Neural Network &MISO& Multiple-Input Single-Output\\
AR &Augmented Reality &NNs& Neural Networks\\
BDMA &Beam-division multiple access &NGMA &Next-Generation Multiple Access\\
BLER& Block Error Rate& NOMA &Non-Orthogonal Multiple Access\\
BnB &Branch-and-Bound &OFDMA& Orthogonal Frequency Division Multiple Access \\
CDMA &Code Division Multiple Access &OMA&Orthogonal Multiple Access\\
CNN &Convolutional Neural Networks &OA-FMTL& Over-The-Air Federated Multi-Task Learning\\
CRN &Cognitive Radio Networks &PDMA&Pattern Division Multiple Access\\
CSAT &Carrier Sense Adaptive Transmission &PHY& Physical\\
%CD-NOMA &Code-Domain NOMA &PD-NOMA &Power-Domain NOMA\\
CSMA &Carrier Sense Multiple Access &PCA& Principal Component Analysis\\
CSMA/CA &CSMA with Collision Avoidance  &QoS &Quality of Service\\
CW &Contention Window& RA& Random Access\\
DCF &Distributed Coordination Function&RFR &Random Forest Regressor\\
DIFS &DCF Inter-Frame space &RS& Rate Splitting\\
DCM& Duty Cycle Management &RSMA& Rate Splitting Multiple Access \\
DT&Decision Tree& RIS& Reconfigurable Intelligent Surface\\
DBN& Deep Belief Networks &RNN& Recurrent Neural Networks\\
DDPG&Deep Deterministic Policy Gradient &RL& Reinforcement Learning\\
DNN &Deep Neural Networks &RB& Resource Block\\
DRL&Deep Reinforcement Learning &6G& Sixth Generation\\
%EDCA&Enhanced Distributed Channel Access (EDCA)&MLO&MultiLink Operation\\
EH &Energy Harvesting &SC &Superposition Coding\\
EM&Expectation-Maximization &SCF& Spectral Correlation Function\\
FL &Federated Learning &SCMA& Sparse-Code Multiple Access\\
1G &First Generation  &SDMA& Space-Division Multiple Access\\
FDD &Frequency Division Duplexing &SDN& Software Defined Networking\\
FDMA &Frequency Division Multiple Access &SIC& Successive Interference Cancellation\\
GSM &Global System for Mobile &SIFS& Short Inter-Frame Spacing\\
HB &Holographic Beamforming &SIM& Spectrum Interference Mitigation\\
IM &Index Modulation &SL& Supervised Learning\\
I/Q &In-Phase/Quadrature& SVM &Support Vector Machine\\
ISAC &Integration of Sensing and Communication &TCMA &Trellis-Coded Multiple Access \\
%ITU &International Telecommunication Union & TDD& Time Division Duplexing\\
IoT &Internet of Things & TDD& Time Division Duplexing\\
KNN &K Nearest Neighbors & TDMA &Time Division Multiple Access \\
LAA  &License Assisted Access  &UAV &Unmanned Aerial Vehicle\\
LTE &Long Term Evolution &URA& Unsourced Random Access\\
LTE-U &LTE in Unlicensed Spectrum &USL& Unsupervised Learning\\
LSTM &Long-Short Term Memory &uRLLC&ultra-Reliable Low-Latency Communication\\
LEO &Low Earth Orbit& V2X &Vehicle-to-Everything\\
LDS &Low-Density Signature & VR& Virtual Reality\\
MAC &Medium Access Control & WANET &Wireless Ad Hoc Networks\\
ML &Machine Learning &WFL &Wireless Federated Learning\\
MR &Mixed Reality &WPT &Wireless Power Transfer\\
MINLP &Mixed-Integer Nonlinear Programming &WSN &Wireless Sensor Networks\\
\shline
\end{tabular}
\label{table_0}
\end{table*}

\section{Introduction}\label{sec1}
\IEEEPARstart{W}{ith} emerging scenarios in the sixth-generation (6G) systems, such as immersive communication, massive communication, hyper-reliable low-latency communication, ubiquitous connectivity, integrated sensing and communication (ISAC), and integrated AI and communication, 6G communication systems are expected to support extremely high data rates (up to 1 Tbps), massive dense connectivity of billions of devices (up to $10^7$ devices per km$^2$), a wide range of coverage ($100\%$ land area and 3D coverage), and high mobility (500-1,000 km/h) with a wide variety of applications and services \cite{letaief2019roadmap}, attracting enormous attention from academia and industry. However, the multiple access (MA) techniques developed in previous generations are not scalable to meet the following unprecedented requirements, such as ubiquitous connectivity, intelligent communication, ultra-high reliability, low latency, integration of network communication and computing, etc. due to limited spectrum resources, high hardware cost, high power consumption, and high complexity. In addition, existing MA schemes face significant challenges, such as ultra-high spectral/energy efficiency to provide ubiquitous connectivity in resource-constrained IoT networks, enabling massive device connectivity with low overhead, providing fairness, achieving strict timeliness for autonomous vehicles and remote surgery services, and enabling compatibility and scalability for 6G and beyond. With the emergence of 6G technologies, resource optimization, interference management, channel estimation, signal processing, and security issues also appeared to be more challenging in MA. These challenges, interacted with the high demands of 6G, are prompting a rethink and redesign of next generation multiple access (NGMA) \cite{liu2022evolution,liu2022developing} techniques for a green and intelligent network.

With billions of smart devices seamlessly connecting to the Internet and a rapidly growing number of bandwidth-hungry mobile broadband devices, it is envisioned that MA and its optimization methods for achieving better Quality-of-Service (QoS), especially for latency-sensitive applications, are attracting more attention. For example, emerging novel applications have posed unprecedented demands on bandwidth, power, latency, and connectivity in wireless networks, including but not limited to the super-smart society (referred to as autonomous driving, smart cities, and smart healthcare), augmented reality (referred to as XR, including augmented reality (AR), mixed reality (MR), and virtual reality (VR)), and automation and manufacturing (Industry 4.0). However, traditional MA design and optimization approaches are losing ground due to a number of drawbacks. For example, low spectral efficiency in the typical orthogonal multiple access (OMA) schemes, spatial division multiple access (SDMA) schemes are not feasible in densely populated urban areas or indoor environments due to the complexity of multi-antenna processing techniques, non-orthogonal multiple access (NOMA) schemes sacrifice their hardware complexity and system overhead to improve overall spectral efficiency, DCF-based random MA schemes are impossible to avoid collisions, etc. Artificial Intelligence (AI) solutions are able to learn the inference information online and make optimal decisions in near real-time. In addition, the intelligence characteristic of 6G has triggered AI applications in MA, and AI-empowered MA designs will exploit the full potential of radio signals to address these requirements and challenges, thereby having been widely investigated by academic and industrial communities.

\begin{figure*}[!htb]
\centering
\includegraphics[width=6.5in]{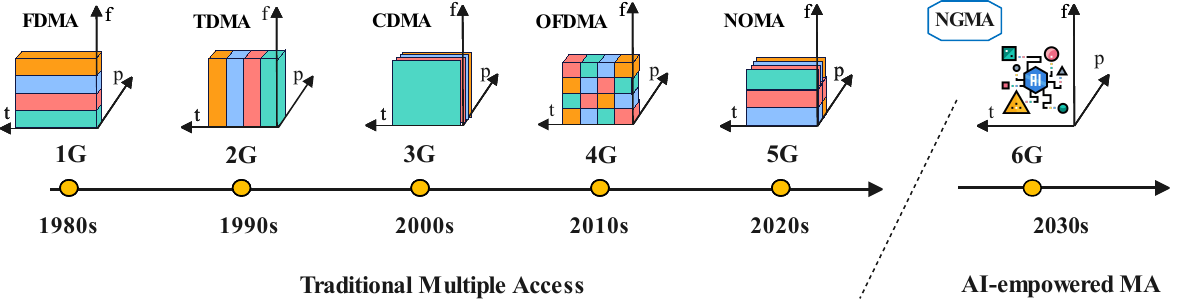}
\caption{Roadmap of multiple access.}
\label{fig_01}
\end{figure*}

\subsection{Roadmap of Multiple Access}
\subsubsection{MA in Cellular Networks}
Over the last few decades, the development of MA schemes has been seen as an important factor in the evolution of cellular networks from the first-generation (1G) to 6G. As shown in Fig. \ref{fig_01}, the MA schemes used in wireless networks, including frequency division multiple access (FDMA) in 1G systems; Time division multiple access (TDMA) \cite{steele1999mobile}, which is widely used in the 2G global system for mobile communications (GSM); Code division multiple access (CDMA)\cite{gilhousen1991capacity}, which supports multiple users with user-specific spreading sequences within the same resource block (RB), was originally proposed by Qualcomm and is widely used in 3G; and orthogonal frequency division multiple access (OFDMA)\cite{li2013ofdma}, which is implemented in 4G networks by dividing the frequency and time RB into narrow subcarriers and slots. A defining feature of all MA schemes is their orthogonality, which ensures that wireless network resources, including frequency, time, and code resources, are theoretically allocated to different users. It allows multiple users to access the network while avoiding multi-user interference. In general, FDMA, TDMA, CDMA, and OFDMA can be regarded as OMA schemes, which typically provide access service to only one user within the same RB at one time. In 4G and 5G, the introduction of multiple-input multiple-output (MIMO) technology creates a spatial dimension, facilitating beam-division multiple access (BDMA)\cite{wang2019multiple} and SDMA\cite{zhang2012turbo}. BDMA and SDMA are capable of serving multiple users on the same RB by utilizing spatial resources and multi-antenna processing. 

\begin{table*}[!htb] 
\newcommand{\tabincell}[2]{\begin{tabular}{@{}#1@{}}#2\end{tabular}}
		%\small
		\centering
			\renewcommand{\arraystretch}{1.2}
			\captionsetup{font={small}} 
			\caption{\scshape Comparison of different MA category} 
			\label{table_1}
			\small
			\centering  
			\begin{tabular}{|m{0.1\textwidth}<{\centering}|m{0.09\textwidth}|m{0.15\textwidth}|m{0.27\textwidth}| m{0.27\textwidth}|}  
				\shline
				%\rowcolor[gray]{0.9}
			    \textbf{MA category}& \textbf{Schemes}&\textbf{Scheduling domain} & \textbf{Advantages} & \textbf{Disadvantages} \\
				\shline  
			   \multirow{4}{*}{\bf OMA} &FDMA & Frequency& $\star$ Simple transceiver design& $\star$ Less number of users are supported\\ 
                        \cline{2-3}
			   \multirow{4}{*}{} & TDMA &Time & $\star$ Avoidance of multi-user interference & $\star$ Low spectral efficiency \\
                        \cline{2-3}
			   \multirow{4}{*}{} & CDMA & Code &$\star$ Easy implementation & $\star$ Inefficient use of spectrum \\
                        \cline{2-3}
			   \multirow{4}{*}{} & OFDMA &Time and frequency& $\star$ Low cost & $\star$ High signalling overhead \\
	              \hline  
			   \multirow{2}{*}{\bf NOMA} & CD-NOMA & Code &$\star$ More number of users are supported & $\star$ Complexity transceiver design \\ 
                        \cline{2-3}
			   \multirow{2}{*}{} & PD-NOMA & Power &$\star$ High spectral efficiency  &  $\star$ SIC error propagation \\
                   \hline  
			   \multirow{1}{*}{\bf RA} & CSMA & Null &$\star$ Highly dynamic and adaptive  & $\star$ High collision and interference \\ 
			    \shline
			\end{tabular}  
	\end{table*}

However, the orthogonality characteristics of the aforementioned OMA schemes limit the number of access users. To allow more users to access the network and significantly improve the overall spectral efficiency, NOMA schemes have been proposed \cite{dai2015non, ding2017application, ding2015impact}. These schemes allow multiple users to share the same slot and frequency RB in the different domains, thus achieving high spectral efficiency at the cost of additional hardware complexity. On this basis, NOMA schemes can be divided into two types: code-domain non-orthogonal multiple access (CD-NOMA) and power-domain non-orthogonal multiple access (PD-NOMA). CD-NOMA separates multiple users via non-orthogonal codes, while PD-NOMA separates multiple users in the power domain. Typical CD-NOMA schemes include trellis-coded multiple access (TCMA) \cite{brannstrom2002iterative}, low-density signature (LDS) sequence-based CDMA \cite{ hoshyar2008novel}, interleave-division multiple access (IDMA) \cite{liu2005analysis}, pattern-division multiple access (PDMA) \cite{chen2016pattern}, and sparse-code multiple access (SCMA). Unlike CD-NOMA schemes, PD-NOMA schemes use superposition coding (SC) techniques at the transmitter and successive interference cancellation (SIC) at the receiver to optimize resource allocation.

To further improve the performance of NOMA, rate-splitting multiple access (RSMA) has recently been proposed with the concept of rate splitting (RS). Specifically, RSMA splits user messages into shared and private content based on RS, then enables the ability to partially decode the interference and partially treat the interference as noise, which is different from the extreme interference management strategies used in SDMA and NOMA. RSMA can dynamically switch between SDMA and NOMA by adjusting power and streams, thereby ensuring effective performance under varying levels of interference \cite{mao2018rate}.

\subsubsection{MA in Wi-Fi Networks}

In contrast, MA schemes in Wi-Fi networks mainly focus on contention-based random access (RA) schemes. The earliest RA schemes, such as ALOHA and slotted ALOHA, limit the channel efficiency to $18\%$ and $36\%$, respectively, \cite{abramson1970aloha,roberts1975aloha}. Carrier sense multiple access (CSMA)-based schemes have been proposed to reduce the collisions suffered from ALOHA-based schemes \cite{kleinrock1975packet}. However, CSMA cannot eliminate collisions due to the hidden and exposed terminal problems. The hidden terminal problem can be solved by using busy tones \cite{tobagi1975packet} or collision avoidance \cite{karn1990maca}. The most classical RA scheme refers to IEEE 802.11, based on CSMA with collision avoidance (CSMA/CA) \cite{ieee1997wireless}, where the distributed coordination function (DCF) is a mechanism to avoid collisions when accessing the channel, its performance has been widely studied \cite{cali2000dynamic, bianchi2000performance, ye2003improving}. The performance of IEEE 802.11 DCF can be influenced by adjusting the contention window (CW) values. In particular, higher CW values result in fewer collisions but increase the number of backoff slots, and lower CW values result in faster backoff but increase the probability of collisions. Therefore, as device density increases, CW values should be set high to avoid collisions, which in turn reduces spectral efficiency. As the number of access devices decreases, the CW values should be set lower to improve spectral efficiency, but this also results in more collisions and long latency. Although IEEE 802.11 DCF performance improvement analysis has been studied, the main problems with RA-based schemes remain, especially for massive connectivity scenarios. In addition, these conventional RA schemes are not well suited for 6G scenarios as the signalling overhead, hardware cost, and complexity will limit their applicability. Table \ref{table_1} compares the different MA in terms of their scheduling domain and features.

\begin{figure*}[!t]
\centering
\includegraphics[width=6.2in]{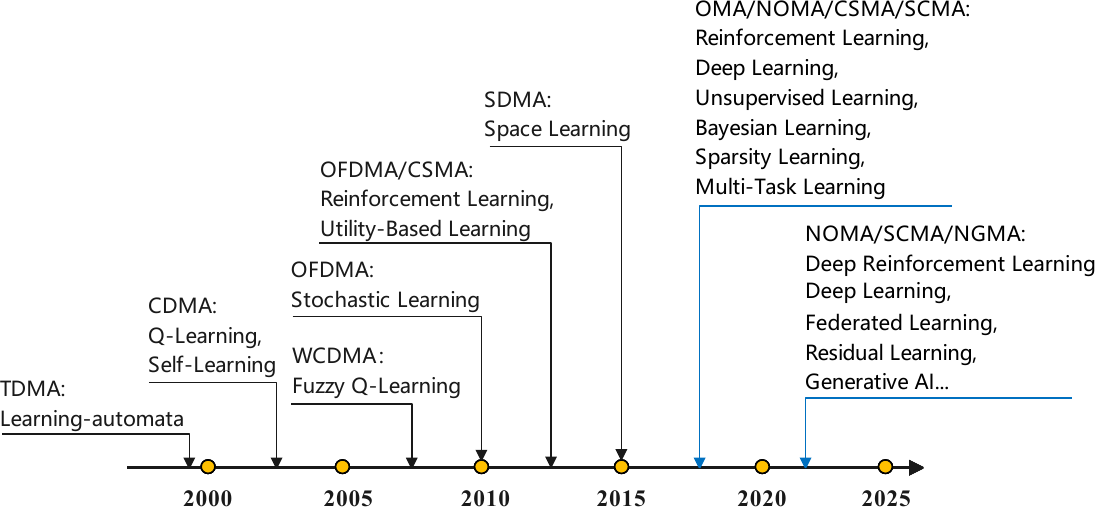}
\caption{Roadmap of AI-empowered multiple access.}
\label{fig_6}
\end{figure*}

\subsection{Evolution of AI for Multiple Access}

Machine learning (ML) was first introduced as a popular technique in the late 1950s and has undergone rapid development. It has been successfully applied to wireless networks and is likely to become a potential technique for significantly improving the performance of next-generation networks \cite{alsheikh2014machine,chen2019artificial,wang2020thirty}. In particular, ML techniques have been used to solve wireless resource management problems, achieving optimal performance in a near-real-time manner, especially in applications such as power control, beamforming, multi-access edge computing (MEC), and reconfigurable intelligent surfaces (RIS). Consequently, the optimization of wireless networks using ML techniques has become a hot topic that has attracted much attention in academia and industry. When considering ML techniques for network architecture, offline and online ML are two fundamental mechanisms; the latter is embedded in network algorithms or protocols, while the former is performed on a local computing unit or can be supported by remote computing units.

Recently, many studies have been conceived on ML paradigms in wireless scenarios, such as HetNets, cognitive radio networks (CRNs), Internet of Things (IoT), wireless ad hoc networks (WANETs), wireless sensor networks (WSNs), etc. In addition, ML techniques have been reviewed in the past decades to address compelling problems of wireless networks \cite{wang2020thirty, xie2018survey, pacheco2018towards, sun2019application, ahmad2020machine, szott2022wi}. Specifically, Wang {\em et al.} \cite{wang2020thirty} addresses the implementation of supervised learning (SL), unsupervised learning (USL), and reinforcement learning (RL) in wireless scenarios. Xie {\em et al.} \cite{xie2018survey} focus on addressing the problems of cyber intrusion detection using ML and data mining. Pacheco {\em et al.} \cite{pacheco2018towards} apply ML algorithms to software-defined networking (SDN) for classification/regression, routing optimization, resource management, etc. Several papers explore the progress of ML techniques in the physical (PHY) layer, medium access control (MAC) layer, network layer, and application layer \cite{zhang2019deep,sun2019application,wang2020thirty,mao2022rate,tang2019future,kurt2021vision,xu2021survey, agarwal2022comprehensive,maraqa2020survey,lopez2022survey,szott2022wi,vaezi2022cellular,liu2022evolution,polese2023coexistence,zheng2023survey, wang2019multiple,shi2019ai,ye2021deep,kim2020two,ahmad2020machine,gao2023grant,liu2022developing,che2023massive,xu2023artificial,fu2023reconfigurable}. More details are shown in Table \ref{table_2}, Compared to these existing surveys and reviews that mainly focus on resource management at the MAC layer and its ML-based solutions, {\em in this paper we mainly focus on the survey of AI-empowered MA in terms of spectrum sensing, protocol design, optimization, applications, and open issues.}

\begin{table*}[!htb] 
\newcommand{\tabincell}[2]{\begin{tabular}{@{}#1@{}}#2\end{tabular}}
		\small
		\centering
			\renewcommand{\arraystretch}{1.1}
			\captionsetup{font={small}} 
			\caption{\scshape Summary of existing survey on AI-empowered MA} 
			\label{table_2}
			\small
			\centering  
			\begin{tabular}{|m{0.06\textwidth}|m{0.04\textwidth}|m{0.08\textwidth}|m{0.15\textwidth}|m{0.5\textwidth}|m{0.03\textwidth}<{\centering}|}  
				\shline
				\rowcolor{mycyan}\textbf{Type} & \textbf{Ref.} & \textbf{Network} & \textbf{Main Scope} & \textbf{Contributions} & \textbf{Year}\\
				\hline 
				\hline 
				\multicolumn{6}{|c|}{\bf Existing} \\
				\hline       
                      \multirow{16}{*}{}& {\cite{zhang2019deep}}  &5G& OMA &  $\star$ Explores the deep learning techniques in the radio access. & 2019\\ 
\cline{2-6}
               	\multirow{16}{*}{}& {\cite{sun2019application}}  &5G & TDMA, OFDMA &  $\star$ Surveys the applications of ML in resource management at the MAC layer, networking and mobility management at the network layer, and localization in the application layers. & 2019\\ 
\cline{2-6}     
                      \multirow{16}{*}{}& \cite{ding2022state}  &6G& NOMA &  $\star$ Surveys the AI applications in RIS-assisted NOMA systems. & 2019\\ 
\cline{2-6}
	                   \multirow{16}{*}{Survey} & {\cite{wang2020thirty}} &  WNs & OFDM, NOMA & $\star$ Reviews ML-enabled wireless networks by elaborating on various learning models. & 2020\\
\cline{2-6}
			  \multirow{16}{*}{}&  {\cite{mao2022rate}}  & 6G & RSMA & $\star$ Presents the first comprehensive tutorial on RSMA by detailing its architecture, taxonomy, applications, and complexities. & 2020\\  
\cline{2-6}
                          \multirow{16}{*}{}& {\cite{tang2019future}} & Vehicular & TDD, OFDMA, NOMA & $\star$ Explores various ML techniques applied to the radio access, networking, and security parts of the 6G vehicular network. & 2020\\ 
\cline{2-6}
                        \multirow{16}{*}{}& {\cite{kurt2021vision}} &B5G or 6G & OFDMA, NOMA & $\star$  Investigates AI-enabled HAPS in terms of design, topology management, handoff, and resource allocation. & 2021\\  
\cline{2-6}
			   \multirow{16}{*}{}&  {\cite{xu2021survey}} &  5G HetNets  & OFDMA, NOMA & $\star$ Investigates the learning-based resource allocation (RA) structures, models, and algorithms. & 2021\\ 
\cline{2-6}
                    \multirow{16}{*}{} & \cite{szott2022wi} & Wi-Fi & RA & $\star$ Surveys IEEE 802.11 performance improvement with machine learning. & 2022\\
\cline{2-6}
			   \multirow{16}{*}{}&  {\cite{agarwal2022comprehensive}} & 5G HetNets &  OFDMA, NOMA & $\star$ Surveys the AI-enabled radio resource management methods for improving spectrum utilization, load balancing, and network energy efficiency. & 2022\\ 
\cline{2-6}
			  \multirow{16}{*}{}&  {\cite{maraqa2020survey}} & B5G or 6G &  PD-NOMA & $\star$ Focuses on optimizing and improving the power domain NOMA with ML. & 2022\\ 
\cline{2-6}
			  \multirow{16}{*}{}& {\cite{lopez2022survey}} &  5G  & Radio Access & $\star$ Surveys the main energy efficiency enabling technologies that 3GPP NR provides massive MIMO, lean carrier design and sleep modes with ML. & 2022\\ 
\cline{2-6}
			  \multirow{16}{*}{}& {\cite{szott2022wi}}  & Wi-Fi & RA &  $\star$ Investigates on improving IEEE 802.11 performance with ML. & 2022\\ 
\cline{2-6}
	             \multirow{16}{*}{}& {\cite{vaezi2022cellular}}  &5G,6G & OMA, RA &  $\star$ Surveys the applications of ML in resource management at the MAC layer, networking and mobility management in the network layer, and localization in the application layer. & 2022\\ 
\cline{2-6}
                  \multirow{16}{*}{}& {\cite{liu2022evolution}}  &6G& NOMA &  $\star$ Explores advanced ML techniques for facilitating the design of NOMA communications. & 2022\\ 
\cline{2-6}
                        \multirow{16}{*}{}& {\cite{polese2023coexistence}}  &6G& NOMA &  $\star$ Explores the spectrum policies and technologies in the above 100 GHz spectrum. & 2023\\ 
\cline{2-6}
  \multirow{14}{*}{}& {\cite{zheng2023survey}}  &Ad hoc& RA &  $\star$ Surveys RL-based MAC protocols and their related solutions. & 2023\\ 
\cline{2-6}
                        \hline
                        \multirow{8}{*}{}&  {\cite{wang2019multiple}} & 5G  & BDMA & $\star$ Investigates the design of multiplexing methods and beamwidth optimization for UAV communications. & 2019\\ 
\cline{2-6}	
                    \multirow{4}{*}{}& \cite{yang2020artificial} & 6G &MAC &  $\star$ Elaborates how to employ the AI techniques to efficiently and effectively optimize the network performance. & 2020\\                         
                  	\cline{2-6}
			   \multirow{8}{*} {Magazine} & {\cite{shi2019ai}} & 5G & NOMA & $\star$ Reviews AI-enhanced spectrum sensing for improving the spectral efficiency of multiple access. & 2020\\
\cline{2-6}
                        \multirow{8}{*}{}& {\cite{ye2021deep}} & MTC &NOMA &  $\star$ Studies NOMA transceiver design with deep learning. & 2021\\  
\cline{2-6}
                        \multirow{8}{*}{}& {\cite{kim2020two}} &  5G &RA&  $\star$ Investigates a framework for estimating access parameters using a DNN model to avoid preamble collision.  & 2021\\ 
\cline{2-6}
                        \multirow{2}{*}{} & \cite{yang2021ai} & 6G & NOMA & $\star$ Reviews AI-driven UAV-NOMA-MEC systems. & 2022\\
                   \cline{2-6}
                        \multirow{8}{*}{}& {\cite{gao2023grant}} & Satellites & NOMA, RA& $\star$ Summarizes the routing technologies from single-layer and multilayer satellite constellations. & 2022\\		
\cline{2-6}	  
                        \multirow{8}{*}{}&  {\cite{liu2022developing}} & 6G & NOMA& $\star$ Applies ML in NOMA and its SIC for massive connectivity. & 2022\\ 
\cline{2-6}
                        \multirow{8}{*}{}&  {\cite{che2023massive}} &  6G  & Unsourced RA & $\star$ Proposes URA and takes federated learning (FL) for privacy protection of URA. & 2023\\ 
\cline{2-6}
			  \multirow{8}{*}{}& \cite{xu2023artificial}  & 5G and 6G &  NOMA & $\star$ Focuses on applying AI technologies in NOMA to achieve automated, adaptive and highly efficient multiple access.  & 2023\\ 	
\cline{2-6}
                       \multirow{8}{*}{}&  \cite{fu2023reconfigurable} & B5G  & NOMA& $\star$ Discusses the potential of ML approaches for RIS-aided NOMA. & 2023\\ 
\cline{2-6}
                       \multirow{8}{*}{}&  \cite{wang2023realizing} &6G  & MD-MA & $\star$ Investigates intelligent multi-dimensional multiple access. & 2023\\ 		   
			    \hline
			    \multicolumn{6}{|>{\columncolor{mygray}}c|}{\bf Ours} \\ 
			    \hline
				\rowcolor{mygray} Survey& {This article} &  6G  & OMA, NOMA, RA & $\star$ Surveys the AI-empowered spectrum sensing, protocol design, optimization, and applications at the MAC layer. & null\\ 
			    \shline

			\end{tabular}  
	\end{table*}

Specifically, the roadmap of ML techniques at the MAC layer is shown in Fig. \ref{fig_6}. In the late 1990s, learning-automata-based TDMA was proposed for stations to select efficiently in each time slot under bursty and correlated traffic \cite{papadimitriou1999self,papadimitriou2000use,papadimitriou2000learning}. Around the year 2000, many ML techniques were proposed for CDMA, such as linear RL was used in CDMA for adaptive energy estimation \cite{chang1999blind}, Tabu learning was used to solve optimization problems of CDMA \cite{li2002tabu}, and self-learning was used to improve the performance of CDMA \cite{liu2005self}. From 2004 to 2009, Q-learning was introduced for radio resource management in multimedia WCDMA \cite{chen2004q}, and fuzzy Q-learning was proposed for admission control in WCDMA \cite{chen2006situation,chen2009fuzzy,glorennec1994fuzzy}. In 2010, distributive stochastic learning was proposed for OFDMA for power and sub-band allocation optimization \cite{cui2010distributive}, and after this, ML has been used in OFDMA for interference management and resource management \cite{bernardo2011intercell,vuvcevic2011reinforcement,fathi2013reinforcement,dai2015game,mertikopoulos2011distributed,wang2013attachment,li2013resource,xu2014distributed}.
Furthermore, RL was used for intercell interference management and resource allocation \cite{bernardo2011intercell,vuvcevic2011reinforcement,fathi2013reinforcement,mertikopoulos2011distributed}, utility-based learning and attachment learning were used for bandwidth allocation \cite{wang2013attachment,li2013resource,xu2014distributed}. In 2015, null space learning was proposed in SDMA for channel interference management \cite{noam2014null}. Meanwhile, ML has also been applied in MIMO-OFDM for resource allocation and optimization \cite{mertikopoulos2016learning, mertikopoulos2015learning} in 2016. Subsequently, DL, RL, and SL approaches have been extensively studied in MA protocols, and especially ML techniques have been explosively studied with the appearance of NOMA and SCMA \cite{xiao2017reinforcement,gui2018deep,cui2018unsupervised,he2019joint, ye2020deepnoma,yang2020cache,huang2020deep,ahsan2021resource,li2020resource,zhong2021ai,ni2022star,zhong2022mobile,lin2023unsupervised,fayaz2021transmit,ni2022integrating}.   In particular, SIC plays an important role on the receiver side of NOMA. The application of AI techniques to reduce the complexity and improve the accuracy of SIC in NOMA needs to be considered for 6G and beyond \cite{xu2023artificial,ding2021no,zhong2021ai}.

In contrast to the above work, the authors of \cite{szott2022wi} investigate the use of ML for layer-specific operation in IEEE 802.11 networks. The new amendments to IEEE 802.11, such as the IEEE 802.11 n/ac/ax/amendments, introduce augmented functionalities (e.g., channel bonding, massive transmissions, short guard interval, and advanced modulations) to improve network operation and user experience. Unlike previous versions, IEEE 802.11be introduces new advanced MA techniques such as OFDMA, multi-resource (MU)-MIMO, multi-AP coordination, and multi-link operation (MLO) to improve spectral efficiency. Compared to IEEE 802.11be, IEEE 802.11bn addresses the latency-sensitive traffic by extending enhanced distributed channel access (EDCA), and also enables multi-AP coordination transmission through coordinated beamforming (CBF) or joint transmission and reception (JTR) \cite{garcia2021ieee,deng2020ieee,wei2024optimized}. However, these augmented functionalities have a limited impact on network performance due to the dynamic nature of Wi-Fi environments and user mobility. By applying ML algorithms to these augmented functionalities, it is possible to gain knowledge, generalize, and learn from experience to design intelligent communication systems. Existing ML solutions for the IEEE 802.11 MAC layer mainly focus on optimizing the parameter design of the DCF. For example, the dynamic selection of CW values based on SL and RL methods can compensate for collisions and idle periods. In addition, self-learning has been used to reduce energy consumption \cite{van2003adaptive,liu2006rl}, support self-configuration \cite{ye2002energy}, and select the MA scheme according to the environmental conditions and application requirements \cite{sha2013self}. In summary, SL \cite{abyaneh2019intelligent,edalat2019dynamically}, RL \cite{zhu2012achieving, amuru2015send, kumar2021adaptive, ali2020performance}, deep reinforcement learning (DRL) \cite{ali2018deep, wydmanski2021contention, zhang2020enhancing}, and federated learning (FL) \cite{ali2021federated, zhang2020enhancing}, models are applied to the IEEE 802. 11 standards \cite{amuru2015send, zhang2020enhancing} and its amendments, most notably in 802.11ac \cite{abyaneh2019intelligent}, 802.11e \cite{zhu2012achieving, coronado2020improvements}, 802.11n \cite{edalat2019dynamically}, and 802.11ax \cite{ali2018deep, wydmanski2021contention}.

Overall, AI technologies are driving MA techniques in terms of resource allocation, optimization, signal processing, sensing, security, dynamic access control, etc. These aspects are driving the latest versions of standards, technical reports, and technical reports and industry white papers. For example, 3GPP Release 17$\&$18 define various advances such as beamforming, MIMO enhancements, mobility enhancements, network power savings, intelligence in MA, IEEE 802.11be and IEEE 802.11bn standards advocate for AI applications to enhance spectral efficiency, enable dynamic access control, facilitate predictive resource allocation, optimize intelligent beamforming, and provide security. Some reports and whitepapers released by IMT-2030, HuaweiTech, or Ericsson propose advanced MA coupled with AI techniques for space-air-ground integrated networks (SAGIN), ISAC, RIS, etc. With the boom of AI technologies, some works on datasets for signal processing, sensing, optimization, and security of MA have been analyzed to improve communication reliability, throughput and latency \cite{yang2020computation,liang2023multi,jiang2019using,alkhateeb2023deepsense,zhou2023aerospace,zhou20246g}. These indicate that the quality of data sets is critical to ensuring the effectiveness and reliability of 6G systems.

\begin{figure*}[!htb]
\centering
\includegraphics[width=7.2in]{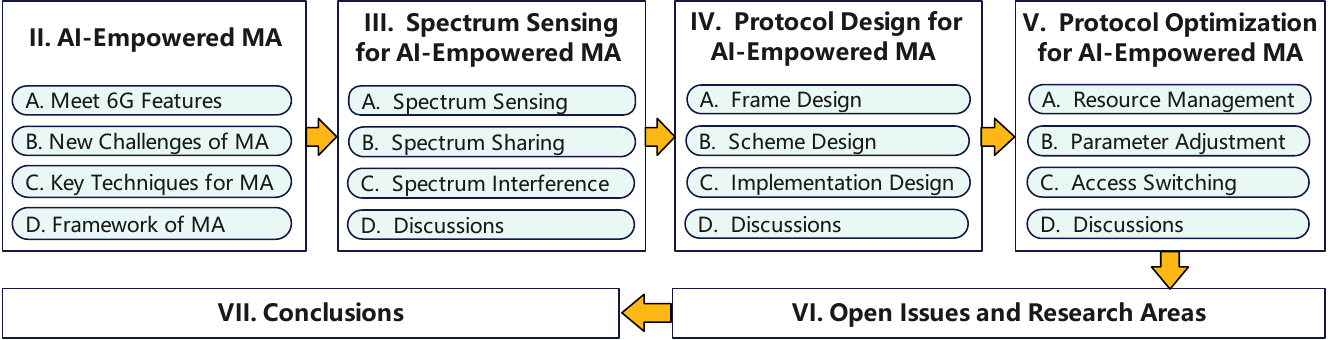}
\caption{Structure of the article.}
\label{fig_1}
\end{figure*}

\subsection{Contributions and Organization}
In this paper, we start with an overview of AI-empowered MA, including the new requirements and key techniques that meet the 6G features, and then propose a unified framework. We continue in the context of spectrum sensing, protocol design, and optimization techniques for AI-empowered MA. Finally, we point out the open issues of AI-empowered MA techniques and discuss the potential applications of AI-empowered MA. Our contributions to this survey are detailed below.
\begin{enumerate}
\item We survey the recent progress and the state of the art in AI-empowered MA techniques. In particular, we first discuss the implementation issues and challenges and then contrast various popular AI-empowered MA approaches to show the recent progress in the interplay between MA and AI techniques in wireless networks. Meanwhile, the basic principles of AI-empowered MA are defined, and the key characteristics are summarized accordingly. By studying the interplay between applied MA techniques and advanced AI techniques in 6G systems, the emerging new requirements, potential techniques, and a unified framework for AI-empowered MA are also illustrated.
\item We elaborate on the spectrum sensing of AI-empowered MA techniques, including spectrum sharing and interference management, which belong to the fundamental problems for improving the performance of AI-empowered MA. We then summarize the AI-empowered MA protocols in terms of the frame, scheme, and implementation designs. Furthermore, we identify the protocol optimization of AI-empowered MA, where efficient AI algorithms are designed for resource management, parameter adjustment, and access switching. The potential problems and solutions based on the current research are also discussed, respectively. 
\item We discuss the open issues and research related to the emerging AI-empowered MA techniques, with particular emphasis on spectrum and interference management, diverse requirements, security and privacy, compatibility, complexity, lower power, and adaptability. In addition, we look forward to applications of AI-empowered MA in typical 6G networks, such as SAGIN, wireless power transfer (WPT), Holographic Beamforming (HBF), Terahertz (THz), ISAC, RIS, MEC, etc.
\end{enumerate}

The remainder of this article is organized as follows (see Fig. \ref{fig_1}). Section \ref{sec2} introduces AI-empowered MA for 6G systems in terms of its emerging features, new requirements, key techniques, and a unified framework model. Section \ref{sec3} presents the possible spectrum sensing, sharing, and interference management techniques for AI-empowered MA that influence spectral efficiency and energy efficiency. Section \ref{sec4} presents the feasible AI-empowered MA protocols including the frame, the scheme, and the implementation designs, and then Section \ref{sec5} explores the optimization of AI-empowered MA protocols in terms of the resource management, the parameter adjustment, and the access switching. Furthermore, Section \ref{sec6} explores the open issues and the research areas of AI-empowered MA that accelerate the evolution of 6G networks. Finally, Section \ref{sec7} concludes this article with an outlook. Table \ref{table_0} gives the list of the abbreviations used here for readability.

\section{AI-Empowered MA}\label{sec2}
AI techniques enable devices to proactively respond to the wireless environment by learning the network states (e.g., channel state information, traffic rates, link, interference, etc.) and link states (e.g., signal-to-noise ratio (SNR), acknowledgment/negative acknowledgment (ACK/NACK), etc.) to achieve a pre-defined goal in a wireless system. As a result, AI techniques have become one of the most important tools and have been widely applied in wireless networks, especially for optimizing and improving the operation of wireless networks. In addition, it is estimated that 5G will reach its capacity limit by 2030, and 6G will emerge to accommodate user services beyond 5G's capabilities, leading to a fully digital and intelligent connectivity era. In this context, AI techniques are expected to play a crucial role in improving the MA performance of 6G services, such as extremely high data rates, massive connectivity, very low latency, high spectral/energy efficiency, high reliability, etc. In this section, the 6G features are aligned with MA in Section \ref{sec2a}, the new challenges of MA are shown in Section \ref{sec2b}, the key techniques enabled on MA are presented in Section \ref{sec2c}, and the framework modules of MA are introduced in Section \ref{sec2d}.

\subsection{Meet 6G Features}\label{sec2a}

\begin{figure*}[!t]
\centering
\subfloat[]{\includegraphics[width=2.2in]{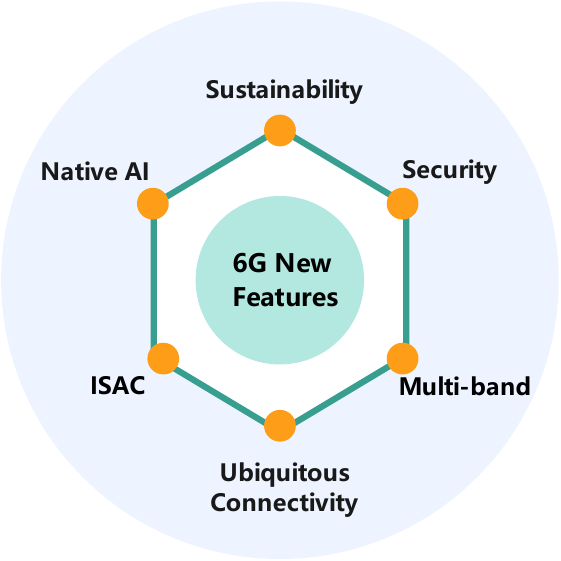}%
\label{fig_first_case}}
\hfil
\subfloat[]{\includegraphics[width=2.2in]{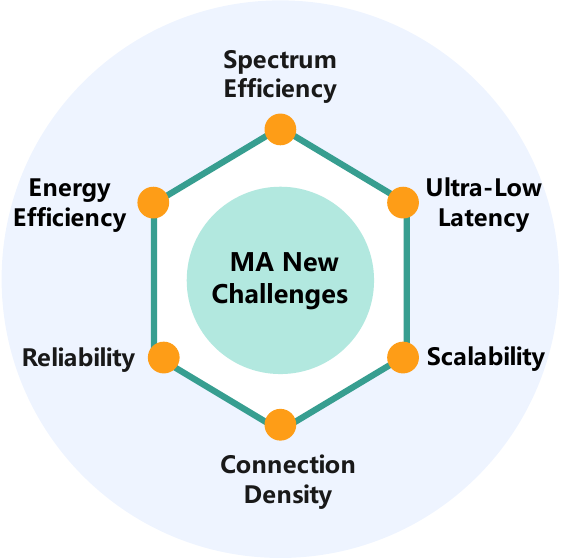}%
\label{fig_second_case}}
\caption{AI-empowered MA meets 6G features and new requirements. (a) Meet 6G features. (b) New challenges.}
\label{fig_6g}
\end{figure*}

According to the 6G trends in IMT-2030 and future network forecasts, the AI-empowered MA technologies align with the 6G features as shown in Fig.~\ref{fig_6g}, which are detailed as follows.

\subsubsection{Native AI}
In contrast to 4G and 5G communication systems with partial or limited AI, intelligence is expected to be a fundamental characteristic of 6G networks, as AI techniques have already penetrated wireless networks and will be ubiquitous across all domains (core, access, edge, device). The wireless evolution from ``connected things" to ``connected intelligence" in 6G will have a profound impact on the evolution of AI and proliferate novel applications. In particular, a new paradigm of AI-empowered MA will simplify and improve radio resource management, adaptability, and efficiency of 6G networks.

\subsubsection{ISAC}
The emergence of ISAC is boosting new services and applications such as automated driving, navigation, etc. In particular, AI-driven ISAC \cite{liu2023ai,jiang2024isac,wang2023generative} in 6G networks will simultaneously improve communication and sensing aspects through full interaction of both, which will help converge the physical, biological, and cyber worlds, making real-time digital twins a reality. To achieve this vision and lay a solid foundation for the intelligence of everything in the future, AI-empowered MA systems are required to satisfy this emerging trend.

\subsubsection{Ubiquitous Connectivity}
With the explosion of low-orbit satellite and UAV technologies in recent years, the integration of ground, air, and space networks in 6G systems will support ubiquitous super-3D connectivity and communications in the vertical dimension. Unlike current terrestrial networks, 3D networks will be able to cover remote areas and provide seamless global access. To make this possible, an efficient AI-empowered MA system must be developed to enable 6G services for users wherever they are.

\subsubsection{Multi-band}
Unlike the limited radio frequency (RF) bands used in 5G, 6G will exploit the rich band resources, such as sub-6 GHz, mmWave bands (30–300 GHz), terahertz band (0.1–10 THz), and even the optical frequency band. It is expected to result in high data rates, with a peak data rate of 1,000 Gbps, and a significant increase in total band capacity by at least 11.11 times. Furthermore, the use of multi-band and massive MIMO technologies is expected to improve spectral efficiency, and the move to higher frequency bands will be a key factor in the investigation of AI-empowered MA for 6G networks.

\subsubsection{Sustainability}
It is estimated that by 2030, research on 6G networks will be in a new era where sustainability will drive its development. This means that 6G wireless networks will not exist in isolation, but should be linked to the triple bottom line of economic sustainability, social sustainability, and environmental sustainability. To realize this sustainable development, a reliable and scalable AI-empowered MA will play an important role in 6G networks.

\subsubsection{Security}
Human-centric 6G network services will require security, confidentiality, and privacy features. The PHY layer security challenges in 5G (e.g., decentralization, transparency, data interoperability, and network privacy vulnerabilities) can be solved by the traditional encryption algorithms, thereby providing secret access and transmission services. However, due to the extensive application of Big Data and AI technologies in wireless networks, the current privacy and security methods in 5G networks will become insecure in 6G networks. Based on this, AI-driven MA security can be ensured by jointly adopting AI security techniques (e.g., adversarial training, federated learning, differential privacy, etc.) and physical security technologies (e.g., encryption, authentication, detection, regular updates, etc.) \cite{wang2022reinforcement,wasilewska2023secure,lv2024safeguarding,xiao2018iot}. Thus, the novel regulation and process of privacy and security for the design of AI-empowered MA has to be considered with respect to AI technologies and MA systems that are still in their early stages.

\subsection{New Challenges of MA} \label{sec2b}
To adapt to the new features and emerging technologies of 6G, AI-empowered MA systems face a number of new challenges.

\subsubsection{Spectral Efficiency}
The increase in the number of connected devices and traffic poses a challenge to improve spectral efficiency, especially the effective use of higher frequency bands (e.g., mmWave and THz) to facilitate the development of 6G communications targeting at Tbps data rate \cite{us2019fcc}. However, in the traditional RA, OMA, and NOMA schemes, the channel access collisions, empty slots, and control overheads can result in low spectral efficiency, thereby degrading the data rate and throughput of the whole system. Meanwhile, the shared spectrum for each device is limited, and it is difficult to accommodate access requests from millions of devices in the network. Therefore, achieving high spectral efficiency becomes a major challenge in the development of MA for 6G networks.

\subsubsection{Energy Efficiency}
The 6G system requires ten times the energy efficiency compared to 5G. Against this background, the interconnection and collaboration of various intelligent entities that have limited or, in some cases, no access to power, the AI-empowered MA paradigm needs to satisfy the green and sustainable development requirements of 6G networks. Consequently, a major challenge for the development of AI-enabled MA is that it can support low-power communications such as on-demand transmission, sleep or wake schedules, and power control to improve energy efficiency. In addition, AI-empowered MA systems are also exploring advanced battery technologies (e.g., energy harvesting (EH), power transfer, etc.) and AI methods to support low-power communications and further eliminate energy waste in 6G networks.

\subsubsection{Ultra-Low Latency and Reliability}
Ultra-low latency and reliable services in 6G systems are important for mission- and time-critical applications (e.g., military, disaster management, etc.). However, with traditional contention-free MAC schemes (e.g., OMA and NOMA), users have to wait for effective resource allocation to grant access to their transmissions, resulting in high latency. Meanwhile, contention-based RA schemes also suffer from latency and reliability issues due to channel access collisions and control overhead. Ultra-low latency and reliability are therefore becoming key challenges in developing AI-empowered MA for 6G networks, especially for ultra-high speed with low latency communication (uHSLLC) services in 6G is a key challenge.

\subsubsection{Massive Connectivity}
The 6G system is expected to connect a massive number of devices (up to 10 million/$km^2$) and provide Gbps coverage everywhere with 3D network coverage (i.e., a global 3D network overlay in ground, air, and space networks), thereby achieving ubiquitous connectivity for exponentially growing devices. The International Telecommunication Union (ITU) expects the number of mobile subscriptions to reach 17.1 billion and the total mobile data traffic to exceed 5TB per month by 2030. Against this background, AI-empowered MA needs to be developed to enable massive, dense, and ubiquitous connectivity for 3D network coverage, which is a major challenge in 6G networks.

\subsubsection{Compatibility and Scalability}
In the context of complex communications in 6G systems, the critical considerations for an AI-empowered MA are compatibility and scalability, which define the ability to cope and perform well with an increasing or expanding workload or scope. In addition, the wireless network environment can be dynamic due to the mobility of the device (e.g., the device joins or leaves the network). Therefore, the challenge is to make the AI-empowered MA protocol scalable and adaptable to the varying wireless network environments without requiring complex computation or additional control overhead while maintaining fairness and access efficiency.

\subsection{Key Techniques for MA}\label{sec2c}
Key technologies such as AI, THz, RIS, Ultra massive MIMO, WPT, HBF, IM, ISAC, and MEC are expected to solve existing MA issues.

\subsubsection{AI}
AI is a fundamental technology in the 6G system \cite{alsheikh2014machine,chen2019artificial,wang2020thirty} because of its ability to solve complex problems without explicit programming. Its successful applications in 6G, such as spectrum sensing, resource management, and MAC layer optimization, have attracted much attention from the industry and research community. In particular, AI technologies have been widely used to solve various problems arising in MAC \cite{szott2022wi,alsheikh2014machine}, such as power allocation, spectrum management, beamformer design, switching control, user association, clustering, scheduling, security, and so on. Compared to the very limited application of AI from 2G to 5G, the full integration of AI and MA in 6G will greatly improve access efficiency with low complexity and low power consumption, especially for ubiquitous intelligent networks.

\subsubsection{THz}
In the 6G system, the near exhaustion of the RF band is pushing the boundaries of the frequency band to the THz band \cite{ghafoor2020mac} to accelerate the deployment of uMUB, uHSLLC, and uHDD services \cite{chowdhury20206g} and enhance the potential of 6G. Unlike the mmWave band, the shorter wavelength THz band can provide hundreds of beams by integrating a large number of antennas \cite{akyildiz2014terahertz,tekbiyik2019terahertz,akyildiz2022terahertz}. THz combined with antenna techniques is possible to increase spectral efficiency and support very high data rates, thereby improving the throughput performance of MA.

\subsubsection{RIS}
RIS is becoming a promising technology that is capable of reconfiguring the wireless propagation environment by exploiting the unique properties of metamaterials-based integrated large arrays of low-cost antennas \cite{huang2019reconfigurable,di2020smart,liu2021reconfigurable,yang2023next}. By doing so, it is possible to increase the coverage area of ultra-small cells, and it also has immense potential to improve the energy efficiency and safety of MA by jointly optimizing RIS size, deployment, phase shift, and transmit power. Furthermore, RIS can also be integrated with existing MAC technology, such as OMA, NOMA, and RA schemes, in order to further improve spectral efficiency while consuming less power \cite{hou2020reconfigurable, liu2022reconfigurable, cao2021ai, cao2022massive, cao2021reconfigurable1, cao2021reconfigurable2}.

\subsubsection{Ultra-Massive MIMO}
Ultra-massive MIMO technology will be critical in 6G systems due to its high spectral and energy efficiency, higher data rates, and higher frequencies \cite{hu2022holographic}. The successful application of ultra-massive MIMO in 6G will trigger the development of uHSLLC, mMTC, and uHDD services. Unlike the conventional MIMO technologies in 5G systems, the ultra-massive MIMO technologies are expected to be integrated with the emerging MA schemes to achieve massive connectivity and high spectral and energy efficiency in the upcoming 6G networks.

\subsubsection{WPT}
WPT offers a new way of powering electric-powered devices, which will be one of the most innovative technologies in 6G communications, breaking the current technical bottlenecks of batteries \cite{zhang2018wireless}. Using WPT technologies in wireless battery charging systems, the electric-powered devices receive wireless power from an electromagnetic field in the air and charge their batteries wirelessly, thus achieving energy migration. In addition, the integration of WPT technologies and MA techniques can support the massive connectivity of power-limited devices, which will be widely applied in some specific regions, such as smart industries and smart cities.

\subsubsection{HBF}

BF \cite{zhang2022holographic,adhikary2023artificial} is a new dynamic beamforming technique that differs from conventional phased array or MIMO systems using a software-defined antenna (SDA). Unlike phased arrays, beamforming is achieved using holograms rather than discrete phase shifters, resulting in symmetrical transmit and receive characteristics for HBF. In 6G, HBF shows its advantages in achieving efficient and flexible transmission and reception of signals in multi-antenna communication devices. In this case, HBF technologies are beneficial for MA systems, including spectrum and energy efficiency, access convergence, and security areas.

\subsubsection{Index Modulation (IM)}
IM is a promising 6G technology for solving existing problems at the MAC layer, such as improving spectral and energy efficiency and reducing complexity. This is because IM uses index patterns as an additional dimension to modulate information bits for data transmission, where the pattern can be the activation state of time slots, subcarriers, transmit or receive antennas, spreading codes, power levels, etc.\cite{li2023index}. MA combined with IM technologies (such as IMMA)\cite{althunibat2019novel} allows each device to transmit additional index bits over partial resources, thereby improving spectral and energy efficiency.

\subsubsection{ISAC}
ISAC is a promising technology for improving spectral efficiency, network reliability, and reducing hardware costs \cite{tan2021integrated}. It will enable 6G systems to continuously sense the dynamically changing wireless environment and establish connections between different devices, thereby supporting autonomous wireless networks. Through ISAC, the full interaction between sensing and communication (i.e., cross-layer, cross-module, and cross-node information sharing) will significantly improve spectral efficiency and mitigate interference. This will enable the mutual benefits from sensing-aided communication and communication-aided sensing, while reducing overall cost and power consumption.

\subsubsection{MEC} 
Due to the limited storage and computing capabilities of devices, large amounts of raw data cannot be efficiently processed by themselves, so MEC technology is proposed to allow devices to offload their tasks to the edge or cloud, thereby accelerating data processing by providing intensive computing capabilities \cite{yang2021joint1,cao2022edge}. On the one hand, highly efficient channel access and data transmission are essential for the implementation of MEC technologies. On the other hand, MA combined with the MEC system can enhance high throughput and timeliness, especially for the upcoming ubiquitous services in 6G scenarios.

\subsection{Framework of MA}\label{sec2d}
To provide further clarification on AI-empowered MA, we present a general framework where the AI approaches and three models with sensing, design, and optimization are detailed.

\subsubsection{General Framework}

\begin{figure}[!t]
\centering
\includegraphics[width=3.5in]{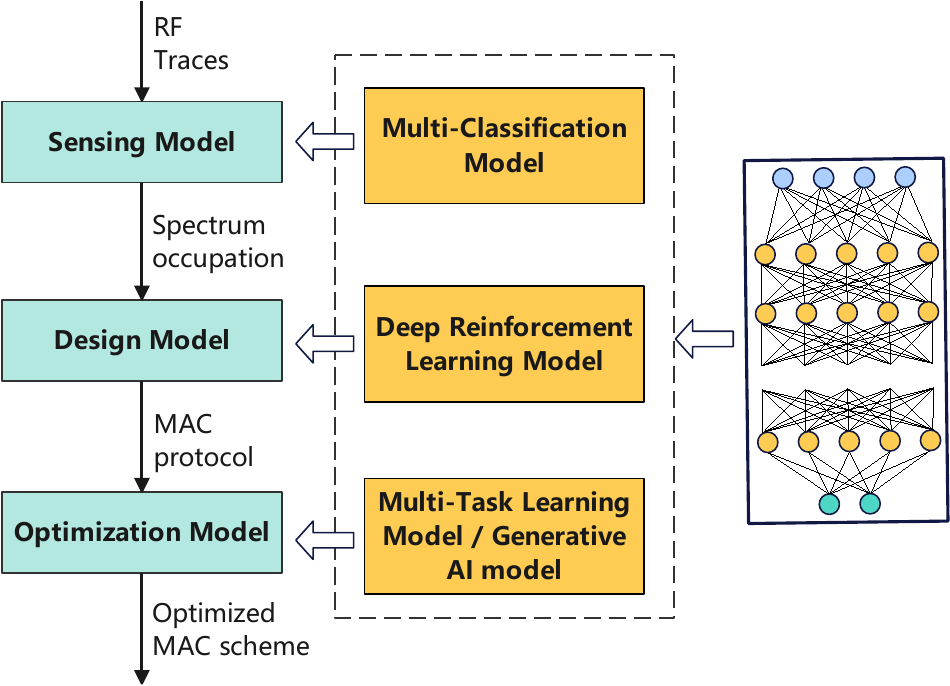}
\caption{AI-empowered MA framework.}
\label{fig_12}
\end{figure}

MA protocols are designed to adapt to the dynamic wireless environment and satisfy various requirements. However, the limited channel resources may restrict the performance of MA protocols when massive users have to access the channel, so the desirable AI-empowered MAC schemes that improve the spectral/energy efficiency while maintaining a low time cost and overhead have to be investigated. On this basis, it is imperative that AI-empowered MA can be flexible, scalable, reliable, and highly efficient as the applications and scenarios change. Fig. \ref{fig_12} shows a general framework of AI-empowered MA schemes, including spectrum sensing, protocol design, and performance optimization, where each module interacts with each other, and all of them can be supported by AI technologies. In particular, accurate spectrum sensing techniques are a prerequisite for MA, which will benefit the design of highly efficient MA protocols by improving spectrum utilization and mitigating spectrum interference. Effective protocol designs are the main body of MA, ensuring effective access and successful transmission. Advanced optimization algorithms are an important part of MA, which can optimize and improve the performance of the designed MA protocols. In addition, AI technologies can be used separately in each part or adopted jointly in three parts to enhance the performance of each part and improve the overall performance of MA for 6G.

\subsubsection{Three Modules}
\begin{figure}[!t]
\centering
\includegraphics[width=3.6in]{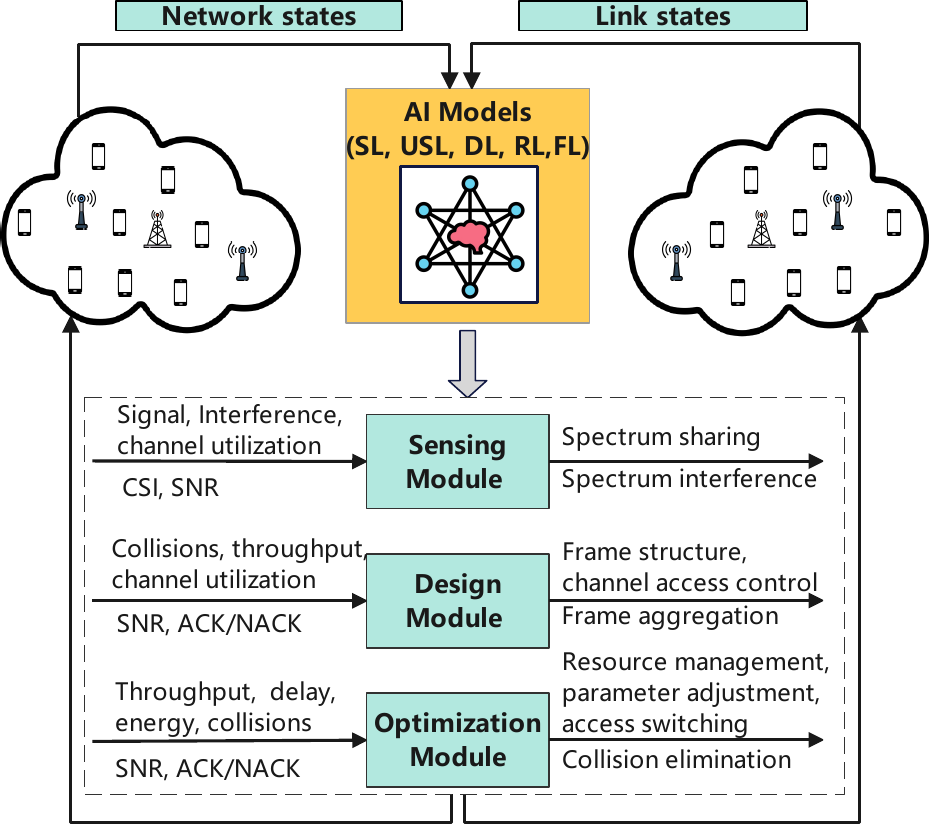}
\caption{Three modules of AI-empowered multiple-access.}
\label{fig_9}
\end{figure}

The implementation of AI-empowered MA protocols refers to three models, i.e., sensing, design, and optimization, the function of each model is shown in Fig. \ref{fig_9}.

\begin{enumerate}
\item Sensing Module: It is a fundamental component for the establishment of AI-empowered MA, especially for its protocol design and optimization. Spectrum sensing is the task of obtaining awareness of spectrum usage to avoid collisions, which not only includes the spectrum usage characteristics across multiple dimensions such as frequency, time, space, code, and beam, but also determines who is occupying the spectrum. Therefore, accurate spectrum sensing can enable the MAC to achieve high spectral efficiency and energy efficiency. In addition, integrating spectrum sensing with AI approaches can achieve better channel coordination and spectrum sharing mechanisms. The sensing model supported by AI technologies is shown in Fig. \ref{fig_9}, where network status information (e.g., signal, interference, channel utilization, link, etc.) is used to achieve accurate spectrum sensing and sharing information with low complexity.

\item Design Module: Protocol design, including frame structure and channel access control, is the most important component for the establishment of AI-enabled MA. It is mainly responsible for coordinating channel access among massive devices in the shared medium using various methods (e.g., centralized, distributed, and hybrid), which affects the performance of the communication system (e.g., network throughput, energy consumption, latency, etc.). In addition, integrating the MA protocol design with AI approaches could further address the diverse service requirements and unique traffic characteristics of devices. The design model supported by AI technologies is shown in Fig. \ref{fig_9}, where network status information (e.g., channel utilization, collisions, throughput, etc.) and link state information are used to support protocol design and frame aggregation, thereby improving channel efficiency with low-cost hardware.

\item Optimization Module: MA protocol optimization refers to resource management (e.g., user association, resource allocation, collision elimination, etc.), parameter adjustment (e.g., CW/backoff setting, frame size setting, transmission rate setting, etc.), and access switching between different MA schemes (e.g., OMA, NOMA, RA, etc.) is essential for obtaining high throughput, low power consumption, and low latency. Additionally, protocol optimization is required to adapt to highly dynamic wireless environments and improve the overall network performance. As shown in Fig. \ref{fig_9}, network status information (e.g., throughput, collisions, delay, power, load, etc.) and link state information are used to support resource management, parameter adjustment, access switching, and collision elimination, thereby optimizing the performance of MA protocols.
\end{enumerate}

\subsubsection{AI Approaches}
We have thoroughly reviewed AI technologies and algorithms for MA systems in UAV, V2X, MEC, and RIS scenarios\cite{yang2021ai,mahmud2021adaptive,yuan2021meta,wang2021interplay,ni2021federated,guo2022distributed}, which are shown in Table \ref{table1_4}. Each approach is characterized as follows.

\begin{table*}[t] 
\newcommand{\tabincell}[2]{\begin{tabular}{@{}#1@{}}#2\end{tabular}}
		\small
		\centering
			\renewcommand{\arraystretch}{1.1}
			\captionsetup{font={small}} 
			\caption{\scshape The AI approaches used in AI-empowered MA schemes} 
			\label{table1_4}
			\footnotesize
			\centering  
			\begin{tabular}{|m{0.16\textwidth}<{\centering}|m{0.24\textwidth}|m{0.35\textwidth}| m{0.15\textwidth}| }  
				\shline
				%\rowcolor[gray]{0.9}
			     \textbf{ML-based Approaches} &\textbf{Suitable Conditions} & \textbf{Applications} &  \textbf{Ref.}\\
				\shline  
			    {\bf SL-based MA} & Solution is unique, a good classic algorithm
 &Spectrum sensing, interference elimination, traffic estimation, channel selection, user association  & \cite{chang2015accuracy, umebayashi2017efficient, thilina2015dccc,zhang2019learning,lin2009machine, amuru2015send, edalat2019dynamically, coronado2020adaptive, coronado2020aios, khastoo2020neura, hassani2021quick}\\ 
	               \hline  
			     {\bf USL-based MA} &  Solution is non-unique, automatic algorithm design & Channel state detection, clustering, frame aggregation, interference cancellation, symbol detection, active user detection, MIMO, RIS configuration, offloading   & \cite{wen2014channel, assra2015approach, zhang2008joint, zhang2017novel, shen2017ica, li2017digital,rooney2021machine}\\ 
                    \hline  
			     {\bf RL-based MA} &  Long-term planning, the action affects the states  &  Resource management, parameter adjustment, interference management, access scheme selection, sensing policy & \cite{sah2022tdma, fathi2013reinforcement, bernardo2011intercell, hassan2021joint, bayat2018multi, tefera2023deep, wang2022deep, zhu2022dynamic, li2022dynamic, huang2022deep, li2020learning}\\ 
			     \hline
                     {\bf DL-based MA} &  Data-driven, is capable of remembering &  Resource allocation, power control, traffic control, user scheduling, interference identification, spectrum access& \cite{singh2022machine, jiang2020multitask, lin2019deep, liu2020situation, camana2022deep, wu2023deep, sharma2020hybrid,miuccio2022energy}\\ 
			     \hline
                     {\bf FL-based MA} &  Proving security and low computing cost  & Security, resource allocation, spectrum sharing, offloading, interference management & \cite{zhong2022over,paul2021accelerated,sery2020analog,wu2021fl,zhang2023efficient,fuhrling2023rate,jeon2022communication}\\ 
	               \shline  
			\end{tabular}  
	\end{table*}

\begin{enumerate}
    \item Supervised Learning: SL algorithms are trained on a certain amount of labeled data set. Given a labeled data set, SL infers a function that maps the input data to the output label. According to the continuity of the output, SL algorithms are extensively used in regression, classification, and prediction, and aim to solve MAC problems in wireless networks, such as spectrum sensing, power control, scheduling, user association, and security issues in UAV, V2X, MEC, and RIS scenarios. The typical SL algorithms mainly refer to support vector machine (SVM) \cite{hearst1998support}, K nearest neighbors (KNN) \cite{beyer1999nearest}, decision tree (DT)\cite{ayodele2010types}, neural networks (NNs) \cite{bengio2009learning}, and Bayesian statistics \cite{rasmussen2006gaussian}.
    \item Unsupervised Learning: USL algorithms try to explore the hidden features of the unlabeled data. Compared to SL, there is no standard accuracy evaluation for the output due to the lack of a labeled data set. USL algorithms aim to classify the data set into different groups by investigating the similarity between them. Generally, USL algorithms are widely used at the MAC layer for solving its clustering and frame aggregation issues in UAV, V2X, MEC, and RIS scenarios. The typical USL algorithms mainly refer to K-Means Clustering \cite{hartigan1979algorithm}, expectation-maximization (EM) \cite{moon1996expectation}, principal component analysis (PCA) \cite{wold1987principal}, and independent component analysis \cite{comon1994independent}.
   \item Reinforcement Learning: RL algorithms allow the agent to learn by interacting with its environment \cite{sutton2018reinforcement}. Specifically, the agent will take the appropriate actions to optimize a long-term objective based on its trial and reward. In contrast to the aforementioned two ML algorithms, RL algorithms are conceived for decision-making relying on their own experience. As expected, RL algorithms have been widely used at the MAC layer for solving mobility management, resource management, and parameter adjustment issues in UAV, V2X, MEC, and RIS scenarios. The most well-known RL algorithms are Q-learning \cite{watkins1992q}, multi-armed bandit learning (MABL) \cite{bubeck2012regret}, actor-critic learning (ACL) \cite{singh2000convergence}, joint utility and strategy estimation based learning (JUSEBL) \cite{perlaza2010can}, and DRL \cite{mnih2015human}.
    \item Deep learning: DL algorithms rely on a multi-layer network for feature extraction and transformation, inspired by the neural network. Specifically, each layer takes the output of the previous layer as input and then infers its output. In principle, some DL algorithms, such as deep neural networks (DNN), deep belief networks (DBN), recurrent neural networks (RNN) with LSTM units \cite{gers2000learning}, and convolutional neural networks (CNN), have been successfully used at the MAC layer to solve resource allocation and optimization problems in UAV, V2X, MEC, and RIS scenarios, thereby improving spectral efficiency and energy efficiency.
    \item Federated Learning: FL algorithms focus on collaborative training between multiple devices, enabling each device to build a general, robust ML model without sharing raw data, and thus becoming the standard for compliance with a range of new regulations on the handling and storage of private data \cite{yang2019federated,yang2022federated}. It is therefore expected to provide reliability, security, and privacy for MA systems while reducing computational complexity issues in UAV, V2X, MEC, and RIS scenarios.
\end{enumerate}

\section{Spectrum Sensing for AI-Empowered MA}\label{sec3}
As mentioned above, the demand for high spectral and energy efficiency at the MAC layer motivates researchers to explore spectrum sensing techniques that support highly efficient channel access. Therefore, spectrum sensing techniques are crucial to improve the performance of MA. Recent advances in MA show that spectral and energy efficiency have been significantly improved by applying AI techniques (e.g., DL, RL, SL, etc. ) in spectrum sensing, spectrum sharing, and interference management. Additionally, AI technologies can potentially be combined with THz or RIS for MA spectrum sensing and dynamic spectrum management in 6G. In this section, AI-enabled spectrum sensing is presented in Section \ref{sec3a}, AI-enabled spectrum sharing is shown in Section \ref{sec3b}, AI-enabled spectrum interference management is depicted in Section \ref{sec3c}, and their discussions are given in Section \ref{sec3d}.

\begin{figure}[!t]
\centering
\includegraphics[width=3.5in]{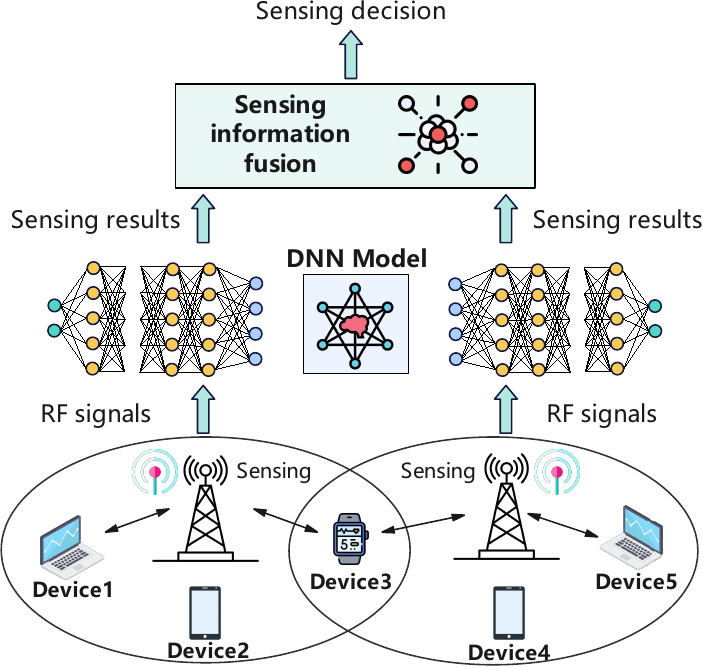}
\caption{AI-empowered spectrum sensing.}
\label{fig_e1}
\end{figure}

\subsection{Spectrum Sensing}\label{sec3a}
Accurate spectrum sensing aims to help devices share the spectrum and access the channel while avoiding multi-user interference. It is considered to be a prerequisite for dynamic channel access and data transmission, and it also poses a challenge for designing high-performance MA protocols, since accurate spectrum sensing depends on the spectrum sensing period. For example, a long spectrum sensing period could reduce the transmission time and consume more energy, thus reducing the spectral efficiency. Therefore, how to realize accurate and effective spectrum sensing has attracted much effort from researchers. In this case, recent studies focusing on AI-enhanced solutions for spectrum sensing and dynamic spectrum management have been widely investigated in cognitive radio communication, which aims to increase spectral efficiency. In addition, THz and RIS combined with AI technologies have also been identified as candidates for spectrum sensing and dynamic spectrum management. A typical AI method applied to spectrum sensing is shown in Fig. \ref{fig_e1}, where the AP senses the CSI and decides the frequency occupation of each device using AI technologies.

Recent improvements in spectrum sensing using various AI techniques for dynamic spectrum management are presented \cite{thilina2013machine,gao2019deep,kulin2018end,song2021deep,li2019survey,shi2019ai,zhou2019dynamic,li2019multi,shah2018reliable,yang2019machine,yang2020improving,yang2021spectrum,yang2021intelligent}. In particular, SL and USL techniques (e.g., SVMs, the K-means clustering algorithm, and the Gaussian mixture model)  are investigated for cooperative spectrum sensing \cite{thilina2013machine,lee2019deep}. Various AI techniques (e.g., SL, USL, QL, DL, etc.) are applied in spectrum sensing and dynamic spectrum management of 6G communications to improve spectral efficiency \cite{zhou2019dynamic}. The deep learning techniques are further applied in CRN for spectrum monitoring, modulation detection, and assignment\cite{li2019survey,gao2019deep,kulin2018end}. K-nearest neighbor learning based reliable spectrum sensing for CNN is investigated to improve spectral efficiency\cite{shah2018reliable}. The multi-agent DRL techniques in resource allocation for D2D communication are presented to improve spectral efficiency where global historical states, actions, and policies are shared without signal interaction\cite{li2019multi}. Q-learning-based approaches are used in cognitive networks to enable secondary users to perform effective spectrum management individually and intelligently, thereby addressing interference coordination between secondary users and interference suppression for primary users \cite{song2021deep}. A long-short term memory (LSTM)-based approach is proposed to extract the energy correlation features and improve the detection performance \cite{xie2020deepl}. A DL method is developed for different types of signal detection based on in-phase/quadrature (I/Q) samples as well \cite{zhang2021signal}, thereby achieving spectrum management. In addition, Saim {\em et al.} \cite{ghafoor2020mac} investigate spectrum management techniques for THz communications, as the new spectrum can be used, it is challenging to implement efficient spectrum management, new radio access, and intelligent automation in THz communications using AI techniques. Yang {\em et al.} \cite{yang2021spectrum,yang2021intelligent} propose intelligent spectrum learning for spectrum utilization and combine it with RIS techniques to significantly improve the spectral efficiency of 6G networks.

Different from AI-assisted spectrum sensing for OMA, Shi {\em et al.} \cite{shi2019ai} investigated learning-based cooperative spectrum sensing for NOMA, which has great potential to improve spectral efficiency and support massive connectivity. An SVM-based SL algorithm is introduced to achieve cooperative spectrum sensing with less prior information requirement, higher sensing accuracy, and lower computational complexity. Compared to traditional OMA mechanisms, SVM-based cooperative spectrum sensing in NOMA allows more than one user to occupy the same time-frequency resource, thus achieving much higher spectral efficiency. However, cooperative spectrum sensing will face more challenges as the number of simultaneous transmissions increases, and the classifiers cannot correct the existing classification error, which affects the accuracy of cooperative spectrum sensing.

\subsection{Spectrum Sharing}\label{sec3b}
Effective spectrum sharing relies on accurate spectrum sensing, which has been extensively investigated as mentioned above. With the development of AI techniques in spectrum sensing, significant efforts have been made to develop intelligent spectrum sharing solutions, which enable AI-enhanced spectrum sharing to be a promising solution due to the reduced need for prior information requirements and higher sensing accuracy. Intelligent spectrum sharing, which improves spectral efficiency, is required to satisfy the following rules: i) interference is reduced or mitigated, or ii) sharing is enabled in time, frequency/band, or spatial domain. Existing research on intelligent spectrum sharing focuses on three main areas: spectrum sharing between secondary and primary users in CR networks, spectrum sharing across licensed and unlicensed bands for coexistence networks (such as Wi-Fi and cellular coexistence), and spectrum sharing between active and passive users \cite{polese2023coexistence}.

\begin{figure}[!t]
\centering
\includegraphics[width=3.5in]{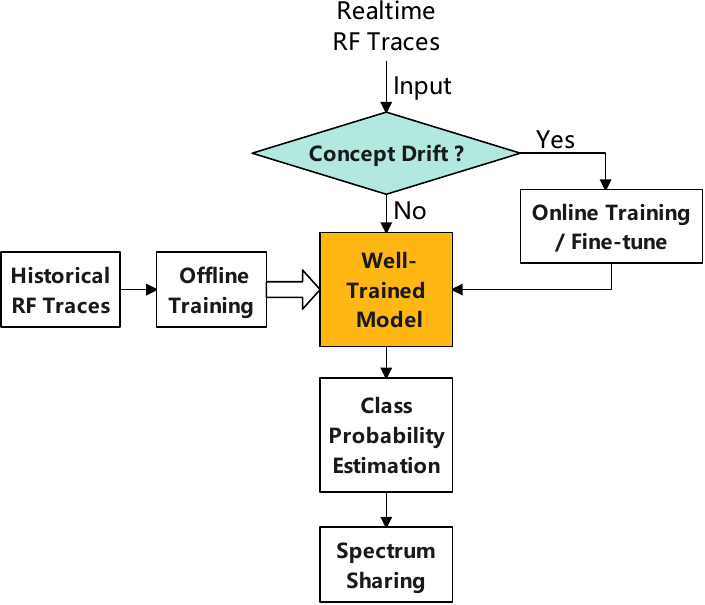}
\caption{AI-empowered spectrum sharing.}
\label{fig_e2}
\end{figure}

For example, the intelligent spectrum sharing solutions for CR networks based on CNN, K-means, Gaussian mixture model, and KNN have been presented \cite{thilina2013machine,lee2019deep, song2021deep}. Baesd on a multi-task learning approach and data sets for resource optimization \cite{yang2020computation}, a great performance improvement has been observed there \cite{liang2023multi}. Fig. \ref{fig_e2} shows a framework for AI-assisted spectrum sharing, where the DL model is trained and tested offline with the historical radio frequency (RF) traces, and the real-time RF traces are examined to determine whether the concept drift has occurred, if not, they are fed to the well-trained model to perform online predictions by signal detection and identification, otherwise they are trained online or fine-tuned and then fed to the well-trained model. Although DL-based spectrum sharing solutions can achieve a better classification performance for ultra-high dimensional models, their computational cost cannot be ignored. Li {\em et al.} \cite{li2018improved} provided an SVM-based spectrum sharing solution, which utilizes the user grouping method to reduce the cooperation overhead and improve the detection performance. Several works use RL-based techniques to control the duty cycle management (DCM) function that is a part of the carrier sense adaptive transmission (CSAT) algorithm or implement the almost blank sub-frame (ABS) allocation mechanism, thereby improving the spectrum sharing efficiency for coexistence networks. For instance, Q-learning is used for adjusting the allocation of LTE subframes in the CSAT duty cycles \cite{naveen2021coexistence}, and it is also used for carrier selection of LTE-LAA systems coexisting with Wi-Fi systems \cite{galanopoulos2016efficient}. Besides, the RL-based spectrum sharing between LTE-LAA and Wi-Fi systems is explored to improve coexistence fairness \cite{han2020reinforcement}. Similarly, the spectrum sharing between active and passive users is explored to alleviate spectrum congestion \cite{bosso2021ultrabroadband}.

\begin{figure*}[!t]
\centering
\includegraphics[width=6in]{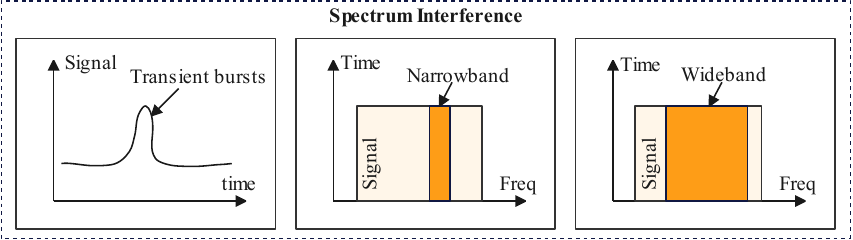}
\caption{Spectrum interference classification.}
\label{fig_SI}
\end{figure*}

\subsection{Spectrum Interference}\label{sec3c}

Spectrum interference or spectrum overlap has a strong impact on sensing performance, and thus spectrum interference mitigation (SIM) becomes a necessary step in sensing processing pipelines to decide how the spectrum should be used. In the context of spectrum sharing, SIM places the burden of interference management on the users, which must be combined for a balanced approach to spectrum sharing. 

In general, spectrum interference is divided into three types as shown in Fig. \ref{fig_SI}, i.e., transient bursts generated by impulse-like waveforms and narrowband or wideband. Spectrum interference can be caused by sources transmitting in the band of interest or by out-of-band emissions, which can lead to the following problems\cite{polese2023coexistence}.
\begin{itemize}
\item Introduces identifiable distortions in the gain or frequency response of the received waveform.
\item Obscures the received signal with a strong and identifiable emission throughout the observed spectrum.
\item Introduces more subtle alterations in the received signals (e.g., when the interference power is close to the noise threshold) that are more difficult to detect and are therefore filtered out.
\end{itemize}

\begin{figure*}[!t]
\centering
\includegraphics[width=4.8in]{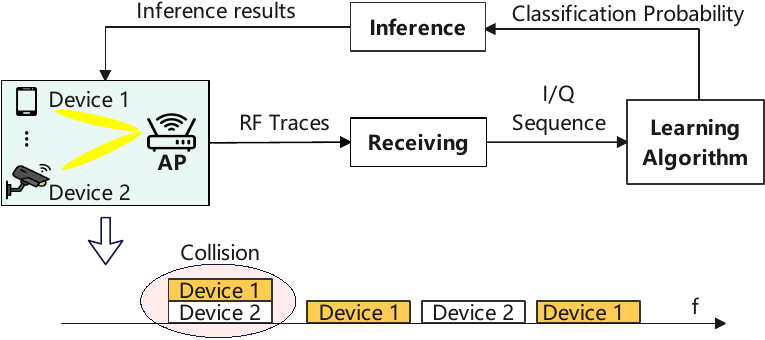}
\caption{AI-empowered spectrum interference management.}
\label{fig_e3}
\end{figure*}

\begin{table*}[t] 
\newcommand{\tabincell}[2]{\begin{tabular}{@{}#1@{}}#2\end{tabular}}
		\small
		\centering
			\renewcommand{\arraystretch}{1.1}
			\captionsetup{font={small}} 
			\caption{\scshape Summary of works on spectrum sensing for AI-empowered MA} 
			\label{table1}
			\footnotesize
			\centering  
			\begin{tabular}{|m{0.04\textwidth}<{\centering}|m{0.14\textwidth}|m{0.06\textwidth}| m{0.07\textwidth}|m{0.27\textwidth}|m{0.15\textwidth}|m{0.1\textwidth}<{\centering}| }  
				\shline
				%\rowcolor[gray]{0.9}
			    \textbf{Ref.} &\textbf{Focuses} & \textbf{ML\  category} & \textbf{ML approach} & \textbf{Applications} & \textbf{Improvements} & \textbf{Scenarios}\\
				\shline  
			  \cite{thilina2013machine}  &Spectrum sharing& DL &CNN  &For cooperative spectrum sensing & Delay &CRN\\ 
 \hline
			   \cite{gao2019deep}&Spectrum sensing &DL  & DNN & For signal detection without prior knowledge  & Spectrum
sensing efficiency  & CRN\\ 
                         \hline 
			   \cite{kulin2018end} &Spectrum sensing& DL & CNN & For spectrum monitoring without complex multi-stage ML & Sensing accuracy & IoT\\ 
                         \hline 
			  \cite{zhou2019dynamic} &Spectrum sensing& RL & DNN& For dynamic spectrum management  & Spectral efficiency & Intelligent IoT\\ 
                         \hline 
			    \cite{song2021deep}& Spectrum sensing& RL  &  DQN  &For dynamic spectrum access without accurate CSI & Achievable data rate and PU protections & CRN\\ 
	              \hline  
			    \cite{xie2020deepl} &Spectrum sensing& DL  & CNN, LSTM & For spectrum sensing detection & Detection probability & CRN\\ 
                        \hline 
			 \cite{yang2021spectrum} &Spectrum sensing& SL & DNN & For exploiting the inherent characteristics of the radio frequency spectrum and help the RIS controller make a decision &Energy efficiency& RIS, 6G\\ 
                        \hline 
			  \cite{li2018improved} &Spectrum sensing& SL & SVM  & For cooperative spectrum sensing with user grouping & Sensing efficiency, security& CRN\\ 
                        \hline
			  \cite{yang2019machine}  &Spectrum sharing& SL & DNN  &For spectrum detection in the unlicensed band & Spectrum utilization & IoT, Wi-Fi \\      
                         \hline 
			  \cite{naveen2021coexistence}  &Spectrum sharing& RL  &  Q-learning  & For spectrum sharing between LTE and Wi-Fi in the unlicensed spectrum & Spectral efficiency & LTE-U, Wi-Fi\\ 
                   \hline  
			    \cite{galanopoulos2016efficient} &Spectrum sharing& RL  &Double Q-learning & For spectrum sharing between LTE-LAA and Wi-Fi in the unlicensed spectrum & Spectral efficiency & LTE-LAA, Wi-Fi\\ 
                        \hline
			  \cite{han2020reinforcement}  &Spectrum sharing& RL & MABL  & For spectrum sharing between LTE-LAA and Wi-Fi with cooperative learning & Access fairness &LTE-LAA, Wi-Fi\\ 
 \hline  
			    \cite{li2018artificial} & Spectrum interference& DL  & ANN    & For spectral interference correction &Prediction error & WN\\ 
                        \hline 
			 \cite{zheng2021spectrum} &Spectrum interference& DL & CNN & For automatic modulation classification with two-level data augmentation&Classification accuracy & CRN\\ 
                        \hline 
			  \cite{van2022deep} &Spectrum interference& DL & DNN  & For replacing the interference cancellation blocks of SIC to facilitate high-integrity detection & SIC performance & WNs\\ 
                         \hline 
			  \cite{yang2021intelligent}  &Spectrum interference& DL  & CNN & For help the RIS controllers infering the interference signals from the incident signals & SINR & RIS\\ 
                        \hline
			  \cite{yang2020improving}  &Spectrum interference& SL & CNN  &For sensing the spectrum to avoid access collisions dynamically & Access efficiency & Wi-Fi\\      
                        \hline  
			  \cite{liu2020decentralized} &Spectrum interference& RL  & LSTM  & For dynamically avoiding mutual interference among automotive radars & Success rate & V2X\\ 
                        \hline
			  \cite{sarikhani2020cooperative}  &Spectrum interference& DRL & CNN &For cooperative spectrum sensing to decrease the signalling  & Spectrum utilization & CRN\\ 
 \hline
			  \cite{liu2020reinforcement}  &Spectrum interference& RL &MABL &For predicting the idle probability of each channel and selecting idle channels & Spectrum access probability & Industry IOT\\ 
			    \shline
			\end{tabular}  
	\end{table*}

SIM solutions involve time, frequency, or spatial filtering, which are based on specific characteristics of the sensed signals. In the time and frequency domains, spectrum interference can be detected and mitigated if it is confined to specific times or sub-bands. Recently, comprehensive SIM technologies based on DL (e.g.,\cite{kulin2018end,tekbiyik2021spectrum,rajendran2018deep,zheng2021spectrum,shi2019deep,li2018artificial,zhang2019deepl,xie2020deepl,van2022deep}) and RL (e.g.,\cite{liu2020decentralized,liu2020reinforcement,sarikhani2020cooperative}) have been investigated to promote spectral efficiency, using DL techniques for spectrum monitoring and signal representation\cite{kulin2018end,rajendran2018deep,li2018artificial}, the modulation classification \cite{rajendran2018deep,shi2019deep,zheng2021spectrum}, the interference identification \cite{zhang2019deepl,tekbiyik2021spectrum}, and SIC \cite{van2022deep}. In particular, Kulin {\em et al.} \cite{kulin2018end} explore the end-to-end learning-based CNN model for spectrum monitoring and signal representations (i.e., temporal IQ data, amplitude/phase, and frequency domain representations that affect the sensing accuracy.). Similarly, Tekbiyik {\em et al.} \cite{tekbiyik2021spectrum} propose a CNN-based spectral correlation function (SCF) to achieve better SIM in cellular bands without a priori information. Li {\em et al.} \cite{li2018artificial} investigate ANN-based DL for SIM, to provide more accurate determinations. Rajendran {\em et al.} \cite{rajendran2018deep} employ LSTM to learn good representations of temporal IQ data, amplitude/phase, and frequency domain representations, aiming to perform automatic modulation classification and achieve SIM. Shi {\em et al.} \cite{shi2019deep} present a DL-based modulation classification for dynamic channel access and SIM in CR networks,  thereby increasing the throughput of secondary users and the success probability of primary users. Zheng {\em et al.} \cite{zheng2021spectrum} use radio signal augmentation in DL to promote modulation classification and SIM. Zhang {\em et al.} \cite{zhang2019deepl} train samples from a 10 MHz band in the 2.4 GHz ISM band and analyze the classification accuracy considering CNN, ResNet, CLDNN, and LSTM, and further reduce the training time by band, SNR, and sample selection while maintaining the classification accuracy. Van {\em et al.} \cite{van2022deep} introduce a DL-based SIC detector to replace the interference cancellation blocks and detect the superimposed symbols for the downlink transmissions of non-orthogonal systems.

Fig. \ref{fig_e3} shows a typical intelligent SIM method to avoid spectrum interference and improve medium access efficiency \cite{yang2019machine, yang2020improving}. A DL algorithm is used to achieve SIM based on the classification probability, i.e., by extracting the I/Q sequences and inputting the I/Q data to the neural network-based predictor, the AP predicts the number of devices from the overlapped signal according to the prediction probability and classifies these devices, so as to avoid spectrum interference. Based on the proposed intelligent spectrum learning, Yang {\em et al.} \cite{yang2021spectrum,yang2021intelligent} further explore the interference mitigation of RIS-assisted communications, where the trained DL model helps RISs to infer the interfering signals directly from the incident signals, and then configures RISs to avoid interference.

Different from DL-based SIM techniques, RL-based SIM techniques are adaptive to the dynamic environment. For example, Liu {\em et al.} \cite{liu2020decentralized} use RL to make a decision in a dynamic environment, to achieve frequency sub-band selection based on the present and past observations. To maximize the spectrum utilization for dynamic CR networks, Sarikhani {\em et al.} \cite{sarikhani2020cooperative} propose a DRL-based cooperative spectrum sensing method to reduce secondary user signalling and perform SIM. Considering the cognitive industrial system, Liu {\em et al.} \cite{liu2020reinforcement} also use DL with Bayesian fusion to dynamically find idle channels while avoiding spectrum interference. Furthermore, by setting double thresholds and weighted energy detection, a better SIM can be achieved in comparison with cooperative spectrum sensing and energy detection.

\vspace{-5pt}
\subsection{Discussions}\label{sec3d}
The developed intelligent spectrum learning technologies for MA protocols have become a popular research area that has attracted much attention from academia and industry. However, AI-empowered spectrum sensing, spectrum sharing, and SIM still face significant challenges.
\begin{enumerate}
\item Security and Privacy: While most of the presented AI-empowered spectrum sensing, spectrum sharing, and SIM for MA functions are centralized, the distributed operation approaches are more suitable for random traffic request scenarios. Moreover, there exist potential security and privacy risks associated with centralized operation, such as sensitive user data being captured at the central controller. Therefore, decentralized and distributed intelligent spectrum learning is encouraged to provide security and privacy, especially the information interaction and sharing are required for multiple agents.
\item Heterogeneity: Intelligent spectrum sharing may face problems of heterogeneity in the underlying technologies, such as different MAC and PHY operations, which are usually managed by separate operators and require either common solutions or coordination between different standardization and technology development forums. In this case, fair sharing issues are therefore still an open challenge.
\item Scalability: Intelligent spectrum learning mostly focuses on the lower frequencies (i.e., below 24 GHz) and active users, so the approaches involved are difficult to apply directly to the ultra-high frequencies (i.e., above 100 GHz) or passive users. Moreover, the combination with other technologies such as THz, RISs, and Massive MIMO complicates the implementation of intelligent spectrum sensing, sharing, and SIM.
\item Power Cost: The cooperative sensing of multiple devices, the combination of different technologies, and the dynamically changing wireless environment result in particularly high power consumption. Therefore, the energy efficiency of intelligent spectrum learning is also a critical challenge.
\end{enumerate}

\begin{figure*}[!t]
\centering
\includegraphics[width=6.8in]{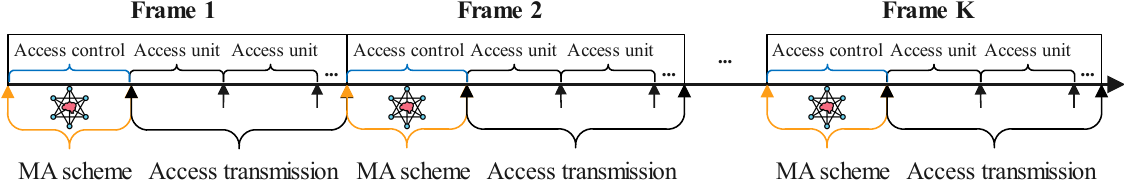}
\caption{The general frame structure of AI-empowered MA.}
\label{fig_f}
\end{figure*}

\begin{figure*}[!t]
\centering
\includegraphics[width=6.8in]{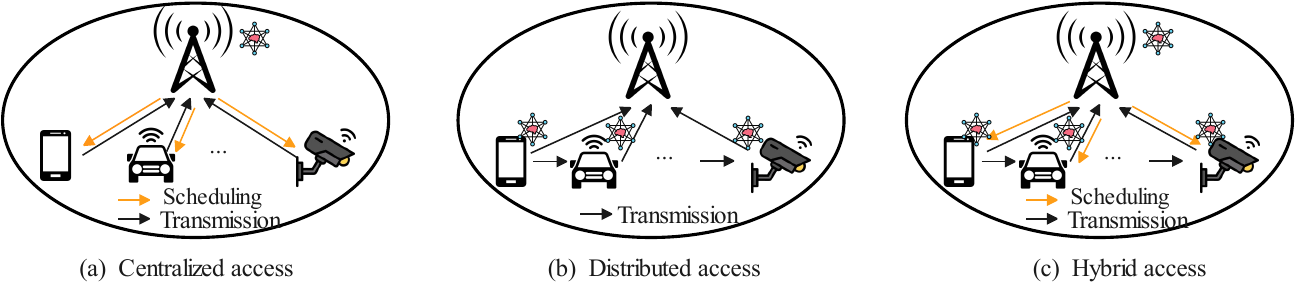}
\caption{Three types of AI-empowered MA: centralized, distributed, and hybrid.}
\label{fig_p}
\end{figure*}

\section{Protocol Design for AI-Empowered MA}\label{sec4}
With the efficiency and accuracy of AI-empowered spectrum sensing, the performance of MA protocols can be further enhanced by AI techniques in each of the three dimensions: collision reduction, throughput improvement, and adaptivity enhancement. Moreover, AI-enabled structure-based designs can significantly enhance the above benefits without introducing additional reliability losses. In addition, AI technologies are expected to be combined with HBF, RIS, or MIMO in MA for 6G ubiquitous intelligent services. In this section, the related designs at the MAC layer include the following three structure-based aspects: frame, protocol, and implementation designs are discussed and summarized. Specifically, the AI-enabled MA frame design is presented in Section \ref{sec4a}, the AI-enabled MA scheme design is shown in Section \ref{sec4b}, the AI-enabled MA implementation is depicted in Section \ref{sec4c}, and their discussions are given in Section \ref{sec4d}.

\subsection{Frame Design}\label{sec4a}

The frame structure is important in the protocol design of AI-empowered MA, and the ability to achieve high MA efficiency is intrinsically related to the designed frame structure. For example, the frame structure design aims to ensure ultra-reliability, fast processing, and wide bandwidth is recommended in \cite{sutton2019enabling}, where a highly flexible frame structure supports resource multiplexing of scheduled users. Pedersen {\em et al.} \cite{pedersen2016flexible} propose a flexible frame structure that integrates frequency division duplexing (FDD) and time division duplexing (TDD). The frame structure allows users to multiplex with dynamically adjusted frame sizes, i.e., the frame size can be dynamically adjusted for each user's scheduling. Pedersen {\em et al.} \cite{pedersen2016system} further explore their flexible frame structure with a dynamically changing frame size. Jiang {\em et al.} \cite{jiang2017state} propose a general frame structure that includes general frames, large frames, and superframes, where the MAC operation is repeated with either a fixed format or a random interval. Liu {\em et al.} \cite{liu2019enhancement} design a superframe structure, where the slot selection algorithm is based on the QL model.

A typical frame structure of various MA protocols is shown in Fig. \ref{fig_f}, which is based on frame division, where each frame is an operation cycle, i.e., a repeated time epoch with either a fixed or a random interval, it is responsible for the confirmation of the MA scheme (e.g., TDMA, CDMA, FDMA, NOMA, RA, etc.) and access transmission, where the MA scheme for accessing channels in an operation cycle can be confirmed using AI techniques, and the then successive access transmission with specific access units depends mainly on the confirmed MA scheme, which can also be modified using AI techniques. The access unit used to share the medium is determined not only by the MA scheme but also by the multiplexing solutions in the PHY layer. Each MA scheme in a frame has different running sequences, exchanged information, capacity, and the number of available access units, which vary according to the protocol design and the function of different MA. Following this general frame structure, the protocol and implementation designs of various AI-empowered MA for their performance enhancement are presented below.

\begin{table*}[!htp] 
\newcommand{\tabincell}[2]{\begin{tabular}{@{}#1@{}}#2\end{tabular}}
		\small
		\centering
			\renewcommand{\arraystretch}{1.1}
			\captionsetup{font={small}} 
			\caption{\scshape AI-empowered centralized protocols: OMA } 
			\label{table_c}
			\footnotesize
			\centering  
			\begin{tabular}{|m{0.05\textwidth}|m{0.11\textwidth}|m{0.1\textwidth}|m{0.35\textwidth}|m{0.26\textwidth}|}  
				\shline
\textbf{Ref.} & \textbf{Protocols} & \textbf{Mechanisms} & \textbf{Applications} & \textbf{Improvements}\\
				\shline                          
\cite{papadimitriou1999self} & TDMA & RL & For operating efficiently under bursty and correlated traffic & Access efficiency\\ 
\hline
\cite{papadimitriou2000use} & TDMA & RL & For the selection of stations in one slot & Real time and spectral efficiency\\ 
\hline
\cite{papadimitriou2000learning} &TDMA &RL& For the selection of transmission in each slot & Delay and network performance\\  
\hline
\cite{fathi2013reinforcement} & TDMA & RL &For adapting MAC parameters with the incoming traffic, buffer, and channel conditions &Energy efficiency and latency\\ 
\hline
\cite{sah2022tdma} &  TDMA  & RL &For optimizing resource utilization & Spectral/energy efficiency\\ 
\hline
\cite{wu2022data} & TDMA & DRL &For data transmission and task calculation& Delay, data rate, and energy consumption\\
\hline
\cite{huang2020throughput} & Grant-free MA & DRL &For mitigating potential pilot sequence collisions& Throughput\\
\hline
\cite{wang2013attachment}  & OFDMA  &RL& For multi-channel allocation strategy & Throughput\\ 
\hline
\cite{bernardo2011intercell} & OFDMA &RL &For frequency resource allocation& Spectral efficiency and inter-cell interference suppression \\ 
\hline
\cite{xu2014distributed} &  OFDMA  &RL & For solving the uncertainty of Nash equilibrium& Layer interference and spectral efficiency\\ 
\hline
\cite{cui2010distributive} &  OFDMA  &DL&  For determining control actions &Delay and energy consumption\\ 
\hline
\cite{singh2022machine} &  OFDMA  &DL & For channel estimation-equalization and carrier frequency offset (CFO) estimation & Spectral efficiency\\ 
\hline
\cite{jiang2020multitask} &  OFDMA  &DL&  For the two tasks-user scheduling and beamforming &Delay and total rate\\ 
\hline
\cite{tefera2023deep} & OFDMA & DRL & For resource optimization & Overall rate performance\\ 
\hline
\cite{wang2022deep} &  OFDMA  &DRL & For beamforming coordination, power and carrier allocation& Sum data rate\\ 
\hline
\cite{zhu2022dynamic} &  OFDMA  &DRL & For allocating time slots, subcarriers, and modulation formats dynamically& Delay and energy efficiency\\ 
\hline
\cite{wu2021fl} & FL-MAC & FL &For MAC privacy and security &Convergence speed and model accuracy\\ 
\hline 
\cite{sery2020analog} & GBMA & FL &For solving the distributed access of MAC & Delay and access efficiency\\ 
\hline
\cite{zhong2022over}  & OA-FMTL MAC  &FL &For solving inter-task interference & Computational complexity\\ 
			    \shline
			\end{tabular}  
	\end{table*}

\subsection{Scheme Design}\label{sec4b}
AI-empowered MAC protocol designs for wireless environments have received considerable attention from academics, consisting of centralized, distributed, and hybrid modes, as shown in Fig. \ref{fig_p}.

\subsubsection{Centralized Schemes}
AI-empowered centralized MAC protocols allow an intelligent central controller to coordinate multiple user access and eliminate collisions using ML techniques (such as DL, RL, etc.). Typical centralized-based AI-enabled MAC protocols include OMA (e.g., FDMA, TDMA, CDMA, and OFDMA) and NOMA schemes. In these schemes, multiple users are scheduled by an intelligent central controller to access the channel over a fixed number of resources (e.g., times, frequencies, codes, and power domains). Although these protocols can avoid collisions by pre-allocating resources, they are not very efficient at low loads and lack flexibility under dynamic network conditions. In particular, TDMA-type protocols have strict time synchronization requirements and result in additional bandwidth and power consumption. In addition, access with a large delay is a concern for delay-sensitive applications. CDMA-type protocols require tight power control to overcome the near-far problem at the receiver due to multiple access interference. Power control imposes computational and hardware requirements that increase the overall system cost. FDMA-type protocols require additional circuitry to communicate and switch between different radio channels at a high cost, and the strict linearity requirement on the medium limits their practicality.

A large number of works applying ML techniques to the design of centralized-based OMA protocols have been studied \cite{wu2022data,wu2021fl,papadimitriou2000learning,sery2020analog,papadimitriou2000use,papadimitriou1999self, sah2022tdma, fathi2013reinforcement, zhong2022over, wang2013attachment, tefera2023deep, wang2022deep, xu2014distributed, cui2010distributive, zhu2022dynamic, singh2022machine, jiang2020multitask, bernardo2011intercell}. To improve spectral efficiency and reduce delay, a learning automata algorithm is applied to TDMA for station or transmission selection in each slot \cite{papadimitriou1999self,papadimitriou2000use,papadimitriou2000learning}. To improve resource utilization, a collaborative multi-agent RL framework for TDMA in wireless sensor networks is proposed \cite{fathi2013reinforcement}, and an RL is used for node scheduling in TDMA \cite{sah2022tdma}, DRL is used in IoT networks to support grant-free MA systems, thus mitigating potential pilot sequence collisions without requiring additional information exchange \cite{huang2020throughput}, and DRL is also combined with TDMA to integrate communication and computing \cite{wu2022data}. Some works focus on improving OFDMA performance using different learning methods \cite{wang2013attachment, xu2014distributed, cui2010distributive, tefera2023deep, wang2022deep, zhu2022dynamic,singh2022machine,jiang2020multitask, bernardo2011intercell}. For example, attachment learning for multi-channel allocation is used to improve throughput \cite{wang2013attachment}, utility-based learning for solving Nash equilibrium uncertainty is used to reduce layer interference \cite{xu2014distributed}, stochastic learning for determining control actions is used to reduce delay and energy consumption \cite{cui2010distributive}, DL for channel estimation and user scheduling to improve spectral efficiency \cite{singh2022machine,jiang2020multitask}, and RL or DRL for resource allocation and beamforming coordination is used to improve sum rate \cite{tefera2023deep, wang2022deep, zhu2022dynamic,bernardo2011intercell}. To provide privacy and security for MAC, FL combined with MAC protocol design is investigated, such as FL-MAC \cite{wu2021fl}, GMBA \cite{sery2020analog}, and \cite{zhong2022over}. Table \ref{table_c} compares the aforementioned AI-empowered OMA protocols in terms of their ML mechanisms, applications, and improvements.

\begin{table*}[!htb]
\newcommand{\tabincell}[2]{\begin{tabular}{@{}#1@{}}#2\end{tabular}}
		\small
		\centering
			\renewcommand{\arraystretch}{1.1}
			\captionsetup{font={small}} 
			\footnotesize
			\centering 
\caption{The comparison of various NOMA, where L, M, and H refer to low, medium, and high, respectively; Sep. and Int. refer to separated and integrated respectively; SC and MC refer to single-carrier and multiple-carrier respectively; SIC, IMD, and MPA refer to successive interference cancellation, iterative multiuser detection, and message passing algorithm, respectively.}
\label{tableNOMA}
\begin{tabular}{|m{0.08\textwidth}|m{0.1\textwidth}|m{0.08\textwidth}<{\centering}|m{0.08\textwidth}<{\centering}|m{0.15\textwidth}|m{0.03\textwidth}<{\centering}|m{0.03\textwidth}<{\centering}|m{0.03\textwidth}<{\centering}|m{0.03\textwidth}<{\centering}|m{0.03\textwidth}<{\centering}|m{0.03\textwidth}<{\centering}|m{0.03\textwidth}<{\centering}|} 
\shline
\multicolumn{2}{|c|}{\multirow{2}{*}{\bf Paradigm}} &{\multirow{2}{*}{\bf Flexibility}}&{\multirow{2}{*}{\bf Complexity}}&{\multirow{2}{*}{\bf Benefits}} & \multicolumn{2}{c|}{\bf Carrier} & \multicolumn{2}{c|}{\bf Modulation}&\multicolumn{3}{c|}{\bf Decoding}\\
\cline{6-12}
\multicolumn{2}{|c|}{} & & &  &\bf SC &\bf MC & \bf Sep. &\bf Int. &\bf  SIC &\bf IMD &\bf MPA \\
\shline
\multirow{8}{*}{\textbf{NOMA}} & PD-NOMA &H& L& Spectral efficiency&\checkmark & &\checkmark & & \checkmark& &  \\
\cline{2-12}
\multirow{8}{*}{} &IDMA &M &L& Diversity & \checkmark& &\checkmark & & &\checkmark & \\
\cline{2-12}

\multirow{8}{*}{} &LDS-CDMA &H & M& Without CSI&\checkmark& &\checkmark & & & &\checkmark \\
\cline{2-12}

%\multirow{8}{*}{} &LPMA & M&M & PD and CD gains &\checkmark & &\checkmark & & \checkmark& & \\
%\cline{2-12}

\multirow{8}{*}{} &LDS-OFDM & L& M& Wideband signals & & \checkmark& &\checkmark & & & \checkmark\\
\cline{2-12}

\multirow{8}{*}{} &SCMA& L& H& Diversity & & \checkmark& &\checkmark & & &\checkmark \\
\cline{2-12}

\multirow{8}{*}{} &PDMA &M &H &Diversity & & \checkmark&\checkmark & & & & \checkmark\\
\cline{2-12}

\multirow{8}{*}{} &RSMA &H &L &Spectral efficiency & & \checkmark& &\checkmark & \checkmark& & \\
\shline
\end{tabular}
\end{table*}

\begin{table*}[!htp] 
\newcommand{\tabincell}[2]{\begin{tabular}{@{}#1@{}}#2\end{tabular}}
		\small
		\centering
			\renewcommand{\arraystretch}{1.1}
			%\captionsetup{font={small}} 
			\caption{\scshape AI-empowered NOMA protocols} 
			\label{tableNOMAC}
			\footnotesize
			\centering  
			\begin{tabular}{|m{0.08\textwidth}|m{0.04\textwidth}|m{0.1\textwidth}| m{0.08\textwidth}|m{0.28\textwidth}|m{0.26\textwidth}| }  
				\shline
				%\rowcolor[gray]{0.9}
			    \textbf{NOMA \ \ \ \  category} &\textbf{Ref.} & \textbf{Protocols} & \textbf{Mechanisms} & \textbf{Applications} & \textbf{Improvements}\\
				\shline  	
			        \multirow{11}{*}{} & \cite{lin2019deep}  & MIMO-NOMA  & DL & For analyzing the channel state information (CSI) and transmission detection  & Detection efficiency, energy consumption\\ 
\cline{2-6}
			  \multirow{11}{*}{}& {\cite{liu2020situation}}  & MIMO-NOMA-OMA &DL & For optimizing multi-dimensional resource allocation & Delay, data rate\\
\cline{2-6}
                      \multirow{11}{*}{}&  {\cite{li2020learning}} &  PDMA   & DRL & For solving optimization problem & Energy consumption\\ 
\cline{2-6}
                      \multirow{11}{*}{}&  {\cite{sharma2020hybrid}} &  HMAS   & DL & For signal detection in a high overloaded multiuser system&Symbol error rate\\ 
\cline{2-6}
			  \multirow{11}{*}{}& {\cite{paul2021accelerated}}  & AGMA  &Gradient learning &For solving distributed access problem&Convergence rate\\ 	 
                        \cline{2-6}	   
                        \multirow{11}{*}{\bf PD-NOMA} & \cite{askari2022q}   & Real-time NOMA & DRL & For scheduling multiple transmissions  &Scalability, delay, energy consumption, throughput \\ 
                        \cline{2-6}
				   \multirow{11}{*}{} & \cite{luo2019deep} & Multi-carrier NOMA &DL &For transmit power optimizing &Energy consumption, data rate \\ 
                        \cline{2-6}
				   \multirow{11}{*}{} & \cite{youssef2021deep} &Downlink NOMA &DRL &For power allocation &Energy consumption, data rate \\ 
                         \cline{2-6}
			   \multirow{11}{*}{} & \cite{amin2023deep} &  NOMA-BF &DRL & For investigating RIS phase shifts design and optimization &Energy consumption, sum rate \\ 
                        \cline{2-6}
			   \multirow{11}{*}{} & \cite{kara2021lightweight} &  NOMA-CRS &Lightweight ML &For the sharing-power allocation &Bit error probability, reliability, energy consumption\\ 
                        \cline{2-6}
				   \multirow{11}{*}{} & \cite{lin2021machine} & NOMA-VLC &DL &For signal demodulator&Scalability, data rate  \\ 
                        \cline{2-6}
				   \multirow{11}{*}{} & \cite{zhang2018machine} &Downlink NOMA & Anderson-Darling ML &For detecting the modulation order of interference signals in downlink NOMA system &Detection accuracy\\ 
                        \cline{2-6}
			         \multirow{11}{*}{} & \cite{zhai2021q} & Downlink NOMA &DRL &For solving resource allocation in multi-antenna downlink NOMA systems &Spectral efficiency, sum rate, energy consumption\\ 
                         \cline{2-6}
				   \multirow{11}{*}{} & \cite{ahsan2022reliable} &Uplink NOMA &DRL &For resource allocation in uplink NOMA-URLLC system&Energy consumption, delay  \\ 
                        \cline{2-6}
				   \multirow{11}{*}{} & \cite{gaballa2023study} &MISO NOMA &DRL &For channel estimation &Sum-rate, estimation accuracy, reliability\\
                        \cline{2-6}
			        \multirow{11}{*}{} & \cite{pan2020deep} &  FTN-NOMA &DL &For sliding-window detection &Detection accuracy, spectral efficiency, delay \\ 
                        \cline{2-6}
			   \multirow{11}{*}{} & \cite{yin2022deep} & Cooperative NOMA &DL &For classifying channels of imperfect CSI and improving the accuracy of CSI &Spectral efficiency, CSI accuracy \\  
                        \cline{2-6}
				   \multirow{11}{*}{} & \cite{elsayed2021deep} &  NOMA-SWIPT &DRL &For efficient data compression in an uplink NOMA communication system  &Energy consumption,compression efficiency\\ 
                        \cline{2-6}
				   \multirow{11}{*}{} & \cite{da2020noma} & Random NOMA &DRL &For allocating RA slots &RA slots allocation efficiency\\    	
\cline{2-6}
			  \multirow{6}{*}{} & {\cite{fuhrling2023rate}} & RSMA & FL & For uplink transmission of edge devices & Delay and energy consumption\\ 
\cline{2-6}
                       \multirow{6}{*}{}& {\cite{camana2022deep}} &  RSMA &DL & For calculating the precoder vector and obtaining the common rate variable & Computational complexity and energy consumption\\ 
\cline{2-6}
                       \multirow{6}{*}{}& {\cite{wu2023deep}} & RSMA  & DL & For improving the robustness against CSI defects & Spectral efficiency and overhead\\		
\cline{2-6}	  
                        \multirow{6}{*}{}&  {\cite{huang2022deep}} & RSMA  & DRL& For power allocation & Energy consumption\\ 
\cline{2-6}
                        \multirow{6}{*}{}&  {\cite{zhang2023efficient}} &  RSMA   & FL &For solving a multi optimization problem  & Delay, reliability\\ 
\cline{2-6}
                      \multirow{6}{*}{}&  {\cite{hassan2021joint}} &  RSMA  & RL & For user clustering of fog radio access networks & Average sum-rate\\ 
\cline{2-6} 
                    \multirow{6}{*}{}&  {\cite{mauricio2018low}} &  SDMA  & DL & For mobile stations clustering and scheduling& Computational  complexity\\ 
\cline{2-6} 
                    \multirow{6}{*}{}&  {\cite{chiang2021machine}} &  SDMA  & RL & For beam tracking and switching in multiple UAVs systems& Transmission efficiency\\ 
			   \hline
				   \multirow{11}{*}{} & \cite{jiang2023residual} &  SCMA &DL &For designing end-to-end autoencoder &Bit error rate, computational complexity \\ 
                        \cline{2-6}
			   \multirow{11}{*}{} & \cite{kim2018deep} &  SCMA &DL & For making the decoding strategy &Bit error rate, computational complexity  \\ 
                        \cline{2-6}
				   \multirow{11}{*}{} & \cite{cheng2021deep} &  SCMA &DL &For multiuser SCMA detection  &Computing complexity, delay \\ 
                        \cline{2-6}
			   \multirow{11}{*}{\bf CD-NOMA} & \cite{bhardwaj2023deep} &  SCMA &DRL &For adaptively constructing codebook &Throughput, reliability, delay  \\ 
                        \cline{2-6}
			   \multirow{11}{*}{} & \cite{miuccio2022flexible}   &  SCMA & DL & For designing encoder and decoder  & Decoding delay, decoding efficiency \\ 
                        \cline{2-6}
			   \multirow{11}{*}{} & \cite{zheng2022knowledge}  & SCMA  & DL & For incorporating prior knowledge into SCMA detection & Delay, data rate\\ 
                        \cline{2-6}
			   \multirow{11}{*}{} & \cite{luo2022novel} &  SCMA &DL &For jointly designing the downlink SCMA encoder and decoder &Error rate, computational complexity  \\ 
                        \cline{2-6}
			   \multirow{11}{*}{} & \cite{su2021scma} &  SCMA &DRL &For selecting the subframe and codebook &Throughput  \\ 
                        \cline{2-6}
				   \multirow{11}{*}{} & \cite{tseng2022cross} & SCMA &DL &For cross-physical-and-application-layer codebook allocation &Delay, scalability, data rate \\ 
        \cline{2-6}
				   \multirow{11}{*}{} & \cite{emir2021deep} & C-NOMA &DL &For cross-physical-and-application-layer codebook allocation &Data rate \\ 
    \cline{2-6}
				   \multirow{11}{*}{} & \cite{han2021deep} & CD-NOMA &DL &For jointly optimizing resource mapping and codebook &Bit error rate\\
			    \shline
			\end{tabular}  
	\end{table*}

Additionally, with the development of various NOMA technologies (as shown in Table \ref{tableNOMA}), extensive research has been conducted on AI-empowered NOMA to further improve NOMA performance, in particular to reduce  the complexity and increase the accuracy of SIC at NOMA receivers. ln conclusion, NOMA performance can be improved by applying AI technologies in to channel estimation, interference mitigation, detection or modulation, resource optimization, and signal processing \cite{xu2023artificial,ding2021no,aref2020deep,zhong2021ai,zhou2021machine,he2023hierarchical}. For example, most AI solutions for various PD-NOMA designs include DL (e.g., \cite{lin2019deep, liu2020situation, camana2022deep, wu2023deep, sharma2020hybrid, luo2019deep, lin2021machine, pan2020deep, yin2022deep}), DRL (e.g., \cite{huang2022deep, li2020learning, askari2022q,youssef2021deep, amin2023deep, zhai2021q, ahsan2022reliable, gaballa2023study, elsayed2021deep, da2020noma}), and FL (e.g., \cite{fuhrling2023rate, zhang2023efficient}). In particular, DL is used in MIMO NOMA for CSI \cite{lin2019deep} and resource allocation \cite{liu2020situation}, in RSMA for the precoder vector \cite{camana2022deep} and robustness \cite{wu2023deep}, in HMAS for signal detection \cite{sharma2020hybrid}, in multi-carrier NOMA for power optimizing \cite{lin2021machine}, in NOMA-VLC for signal demodulation  \cite{lin2021machine}, in FTN-NOMA for sliding window detection \cite{pan2020deep}, and in cooperative NOMA for channel classification \cite{yin2022deep}. DRL is used in RSMA and PDMA for resource allocation and optimization \cite{huang2022deep,li2020learning}, in up/downlink NOMA system for transmission scheduling \cite{askari2022q}, resource allocation \cite{youssef2021deep,zhai2021q,ahsan2022reliable}, beamforming \cite{amin2023deep}, channel estimation \cite{gaballa2023study}, and data compression \cite{elsayed2021deep}, and in random NOMA for slot allocation\cite{da2020noma}. FL is used in RSMA for optimizing \cite {zhang2023efficient} and edge transmission \cite{fuhrling2023rate}. On the other hand, some works aim at implementing DL (e.g., \cite{jiang2023residual, kim2018deep, cheng2021deep, miuccio2022flexible, zheng2022knowledge, luo2022novel, tseng2022cross, emir2021deep, han2021deep}) and DRL (e.g., \cite{bhardwaj2023deep,su2021scma}) approaches on CD-NOMA. For example, DL is used in SCMA for encoder and decoder \cite{jiang2023residual,kim2018deep,miuccio2022flexible,luo2022novel}, detection \cite{cheng2021deep,zheng2022knowledge}, and codebook allocation \cite{tseng2022cross}, and also in C-NOMA or CD-NOMA for codebook allocation \cite{emir2021deep, han2021deep}. DRL is used in SCMA for codebook construction and selection \cite{bhardwaj2023deep,su2021scma}. ML algorithms are used to address MA optimization in SAGIN \cite{zhou2021machine,he2023hierarchical}. Table \ref{tableNOMAC} summarizes the typical AI-empowered NOMA in terms of their ML mechanisms, applications, and improvements.

\subsubsection{Distributed Schemes}
In contrast to the centralized schemes, AI-empowered distributed MA protocols allow each intelligent user to compete for multiple accesses and avoid collisions. In this case, scheduling by the BS or AP is not required, and each intelligent user accesses the channel using DL or RL approaches in CSMA-based schemes. Typical RA protocols are the IEEE 802.11 channel access mechanism known as carrier sense (i.e., DCF) and some derivative schemes based on CSMA/CA. Such protocols are slotted, and the busy/idle state of channels is determined by carrier sensing; they perform well in a small network and can eliminate collisions to some extent while satisfying users' RA requirements. However, as user density increases, their delay and throughput degrade rapidly, especially as the load approaches saturation. They also fail to scale as the network expands, wasting additional energy may occur due to collisions, idle listening, and information exchange overheads.

\begin{figure}[!t]
\centering
\includegraphics[width=3.5in]{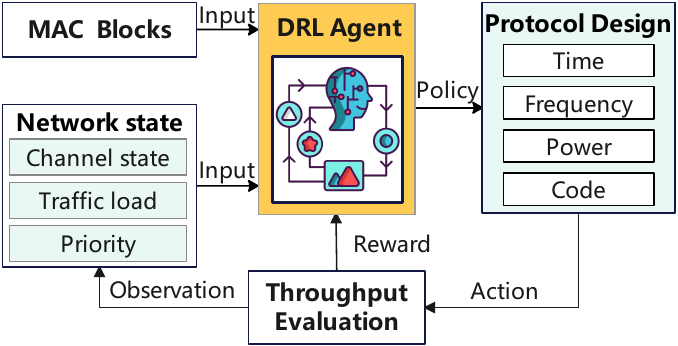}
\caption{MA protocol design with RL.}
\label{fig_des}
\end{figure}

Different from the centralized MAC protocols, there are many AI solutions for the IEEE 802.11 MAC layer to adaptively design the internal parameters of the core Wi-Fi functions. Typically, these AI solutions for distributed MAC protocols include CW value selection \cite{ ali2018deep, coronado2020improvements,kumar2021adaptive, abyaneh2019intelligent, edalat2019dynamically, wydmanski2021contention}, backoff value selection \cite{zhu2012achieving, zhang2020enhancing, amuru2015send, ali2020performance}, time slot selection \cite{lee2020collision,kihira2020adversarial,guo2022multi}, frame size selection \cite{lin2009machine, coronado2020adaptive, coronado2020aios, khastoo2020neura, karmakar2019online}, transmission rate selection \cite{khastoo2020neura, chen2021experience}, and link parameter configuration \cite{moon2021neuro, ali2021federated, hassani2021quick, karmakar2020deep}. In particular, RL-based and DRL-based approaches are generally adopted to conceive distributed MA protocols and set channel access parameters, and SL-based approaches are used to predict CSI. In the proposed RL and SL models, loss functions and rewards are addressed in terms of reduced collisions \cite{ali2018deep, kumar2021adaptive}, the increased difference between successful and collided frames \cite{zhu2012achieving}, improved channel utilization \cite{abyaneh2019intelligent}, increased successful channel access attempts \cite{edalat2019dynamically, zhang2020enhancing}, throughput \cite{wydmanski2021contention}, network utility \cite{moon2021neuro}, and a combination of improved throughput, reduced energy, and reduced number of collisions \cite{amuru2015send}. We provide a summary of the most important ones in Table \ref{tableRA}, while an illustrative example of using RL to optimize MAC protocol design is given in Fig. \ref{fig_des}, where an agent learns a model based on the current MAC building blocks and network configuration to evaluate how the environment affects the MA protocol design from their interaction. The agent then uses the learned model to adaptively design an MA protocol that includes the selection of CW/backoff values, time slots, frame sizes, and transmission rates. 

\begin{table*}[!htp] 
\newcommand{\tabincell}[2]{\begin{tabular}{@{}#1@{}}#2\end{tabular}}
		\small
		\centering
                \renewcommand{\thetable}{IX}
			\renewcommand{\arraystretch}{1.1}
			\captionsetup{font={small}} 
			\caption{\scshape Distributed AI-empowered MA protocols } 
			\label{tableRA}
			\footnotesize
			\centering  
			\begin{tabular}{|m{0.05\textwidth}|m{0.18\textwidth}|m{0.1\textwidth}|m{0.3\textwidth}|m{0.26\textwidth}|}  
				\shline
				\textbf{Ref.} & \textbf{Protocols} & \textbf{Mechanisms} & \textbf{Applications} & \textbf{Improvements}\\
				\shline    
\cite{ali2018deep} & IEEE 802.11ax &DRL & For CW selection & Throughput, delay, and fairness \\ 
\hline
 \cite{kumar2021adaptive} & IEEE 802.11 standard  &RL & For minimum CW parameter selection & Network-level utility\\ 
\hline                     
\cite{zhu2012achieving} & IEEE 802.11e & RL & For backoff value selection & Flexibility and overhead \\
\hline
\cite{abyaneh2019intelligent} &IEEE 802.11 standard & SL &For the minimum CW value selection & Fairness\\  
\hline
\cite{edalat2019dynamically} & IEEE 802.11n  & SL & For CW value selection  & Average throughput, delay, and fairness\\ 
\hline
\cite{zhang2020enhancing} & IEEE 802.11 standard & DRL  & For back-off value selection  & Throughput and fairness\\		
\hline
\cite{wydmanski2021contention} & IEEE 802.11ax & DRL & For CW values selection  & Efficiency and computational cost\\ 
\hline
\cite{moon2021neuro} & IEEE 802.11 standard & DRL  & For interference patterns and configurations  & Delay\\
\hline
\cite{amuru2015send} &  IEEE 802.11 standard  &RL & For back-off value selection  & Throughput and convergence speed\\ 
\hline
\cite{ali2021federated} & IEEE  802.11ax  & FL &  For channel access parameter selection & Throughput\\ 
\hline
\cite{coronado2020improvements} & IEEE 802.11 standard & SL &For AIFS and CW selection & Throughput and QoS\\ 
\hline
\cite{lin2009machine} & IEEE 802.11 standard  & SL & For frame-size selection  & Effectiveness and reaction speed\\ 
\hline
\cite{coronado2020adaptive} & IEEE 802.11 standard & SL  &For frame size selection &Goodput\\ 
\hline
\cite{coronado2020aios}& IEEE 802.11 standard & SL &For frame size selection& Aggregated network throughput \\ 
\hline
\cite{khastoo2020neura} & IEEE 802.11 standard   & SL & For frame size and transmission rate selection& Throughput\\ 
\hline
\cite{hassani2021quick} & IEEE 802.11ac & DL &For solving the backhaul& High rate and low delay \\ 
\hline	                      
\cite{lee2020collision} & IEEE 802.11 standard & RL & For backoff value selection & Collisions and network performance\\ 
\hline
\cite{ali2020performance} &  IEEE 802.11 standard   & RL & For backoff value selection & QoS\\ 
\hline
\cite{karmakar2019online} & IEEE 802.11ac & RL &For frame size selection & Energy efficiency\\ 
\hline
\cite{chen2021experience}  & IEEE 802.11ac & DRL & For experience-driven rate adaptation & Overhead\\
\hline 
\cite{karmakar2020deep} & IEEE 802.11ac  & SL & For link configuration parameter selection  & Link layer performance \\ 
\hline
\cite{guo2022multi} &  IEEE 802.11 standard   &RL & For time slot selection & Delay, robust, and efficiency\\ 
\hline
\cite{kihira2020adversarial} & IEEE 802.11be & RL & For time slot selection & Interference and sum throughputs\\ 
\hline
\cite{szott2022wi} & IEEE 802.11bn  & ML & Resource allocation & Throughput, delay\\ 
\shline
			\end{tabular}  
	\end{table*}

\subsubsection{Hybrid Schemes}
AI-empowered distributed MA protocols easily adapt to dynamic networks and are more suitable for small and low load networks. AI-empowered centralized MA protocols eliminate collisions and achieve high channel utilization at higher loads. AI-empowered hybrid MAC protocols combine and balance aspects of both, where users can be scheduled to access the channel by an intelligent central controller, and the intelligent users can also access the channel via the contention scheme. Switching between RA and scheduled access effectively reduces collisions of RA protocols at high loads and improves channel utilization of scheduled access at low loads. However, collisions during the random access, which relies on slot/code/frequency/power resource reservation of hybrid protocols, become the bottleneck preventing the network from achieving high utilization. In addition, dynamic switching usually results in wasted slots, frequencies, and additional overhead. 

In general, existing hybrid MA protocols combine CSMA with TDMA, FDMA, and CDMA and can achieve dynamic switching between two modes, these protocols behave as the distributed access (e.g., CSMA) at low contention levels and switch to the centralized access (e.g., TDMA, FDMA, or CDMA) at high contention levels \cite{rhee2005z, salajegheh2007hymac, zhang2010hybrid}, or they allocate slots and codes according to CSMA-based bandwidth requests\cite{shu2009energy}. To overcome the above bottleneck and further enhance the benefits of both, various AI techniques are used to achieve the switching between the distributed mode and the centralized mode, thus satisfying access flexibility while maintaining high access efficiency. For example, Qiao {\em et al.} \cite{qiao2018intelligent} propose an intelligent MAC switching scheme to support large and dense networks with low computational complexity, where the contention MAC protocol combined with contention-free MAC protocol is presented, the dynamic switching of both based on ML techniques can adapt to the changing network state. Yang {\em et al.} \cite{yang2018scalable} design a scalable intelligent MAC protocol that combines the contention access and the scheduling access to improve the system throughput, where the time length of the contention access in each superframe is selected based on the learned network states. Yang {\em et al.} \cite{yang2019machine} further optimize the proposed scalable intelligent MAC protocol for coexisting networks including Wi-Fi devices and IoT devices. Cao {\em et al.} \cite{cao2019deep} investigate the coexisting networks consisting of Wi-Fi devices and backscatter devices, and propose a hybrid MAC protocol that combines the contention access with the reservation access using DRL. In addition, Cao {\em et al.} \cite{cao2021ai} develop an AI-assisted hybrid MAC protocol for RIS-based multi-user communication systems, where AI techniques are used to jointly design centralized access and distributed access in a superframe. Gomes {\em et al.} \cite{gomes2020automatic} use RL techniques to switch the MAC protocol between the contention scheme and the reservation scheme based on the current network demand. In addition, several hybrid MAC protocols using RL  techniques are proposed to achieve the scheme selection between OMA and NOMA to adapt to the different requirements \cite{sharma2020hybrid,chaieb2022deep}.

\subsection{Implementation Design}\label{sec4c}
\begin{figure*}[!htb]
\centering
\includegraphics[width=6.8in]{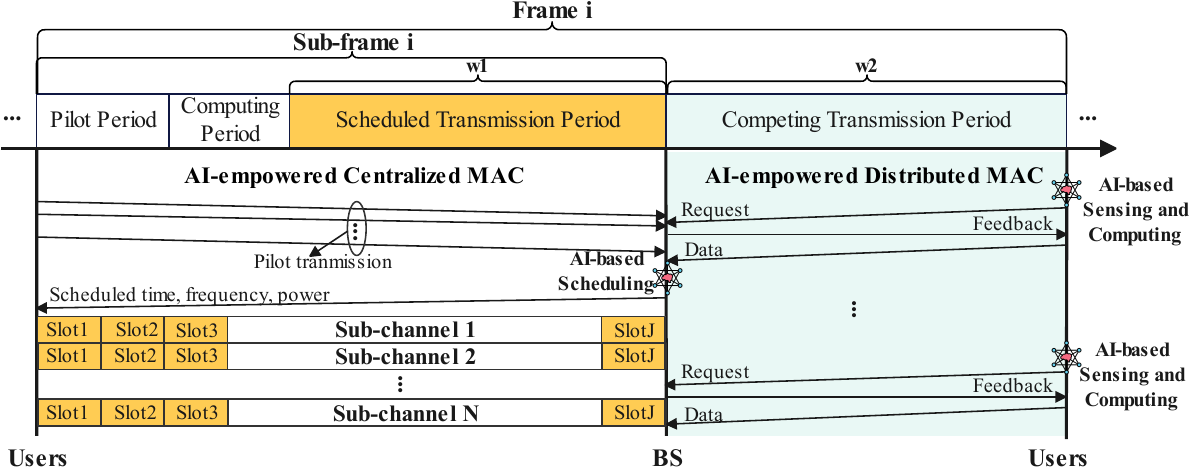}
\caption{Implementation design.}
\label{figh}
\end{figure*}

Referring to these discussed intelligent MA scheme designs \cite{cao2021ai}, a typical implementation design of AI-empowered centralized, distributed, and hybrid MA framework is shown in Fig. \ref{figh}.

In the proposed AI-empowered centralized MA framework, the center controller (e.g., BS or AP) tightly coordinates the multiple access of users. Explicitly, the center controller schedules and allocates resources for users via AI models, as detailed in the left part of Fig. \ref{figh}, each subframe is divided into three periods: the pilot period, the computation period, and the scheduled transmission period. The pilot period and the scheduled transmission period can be further divided into $K$ pilot slots and $J$ data slots, respectively. The users transmit their data in the $J$ data slots on the $N$ non-overlapping subchannels. Based on this frame structure, the AI-empowered centralized MA protocol is designed with full consideration of channel access, computation, and data transmission. The AI-empowered centralized MA protocol combines with TDMA and FDMA, where each user follows the time division scheme in each subchannel. In particular, after synchronization, each user initiates pilot transmission to the central controller in dedicated pilot slots. During the computation period, the central controller first estimates the CSI, followed by time, frequency, and power resource allocation. The central controller then configures the MA parameters and schedules the access of users requiring their data transmissions. Given the excessive complexity of high-dimensional, full-search-based centralized MA protocols, a DL-based computational model is trained offline at the central controller and can be used to find a near-optimal solution with reduced complexity. The input to the trained DL model can be the number of users, the number of subframes, the number of channels, and the channel information. Online inference, which shifts the complexity to offline training, is performed at the central controller to determine the MA parameters and the resource allocation strategy. More explicitly, these related learning tasks share the same input parameters, so that learning multiple related tasks together can improve the prediction accuracy and generalization ability compared to learning them separately.

In contrast to the centralized scheme, in the proposed AI-empowered MA framework, each user configures multiple access and calculates the required resources based on the dynamic network environment. In this case, no assistance from the central controller is required. The users must negotiate with the central controller for channel access and perform the subsequent data transmission, as shown in the right part of Fig. \ref{figh}. In particular, channel sensing and computation are carried out at the user side via RL to determine the MAC parameters and access the channel. The AI-empowered distributed MA protocol combines CSMA and FDMA, where each user follows the IEEE 802.11 DCF scheme in each channel. Specifically, a competing user senses the state of each subchannel. Once a channel is sensed to be idle, the user contends for access to the channel. Waiting for a DCF inter-frame space (DIFS) and backoff, the user sends a request to the central controller. If the channel is available to the user, after a short inter-frame spacing (SIFS) the central controller sends its feedback to the user. After a SIFS, the user then transmits the data to the central controller. Here, an RL-based calculation model can be used by each user to set MA parameters (e.g., CW value, backoff/DIFS value, frame size). The RL model includes the current MA parameter setting and the current network configuration. Model actions include updating the CW value, updating the backoff/DIFS value, and updating the frame size. The reward function is determined by the user's throughput requirement. If the action taken by the user increases the throughput, the user will receive a positive reward. Correspondingly, if the action taken by the user reduces the throughput, the user will receive a negative reward.

\begin{table*}[!htp]
\centering
\caption{The comparison of AI-empowered centralized, distributed, and hybrid MA schemes.}
\small
\label{tableC}
\begin{tabular}{|c|c|c|c|c|c|c|c|c|c|c|c|c|c|c|c|}
\shline
\multicolumn{2}{|c|}{\multirow{2}{*}{Ref.}}  & \multicolumn{4}{c|}{\bf MA schemes}&\multicolumn{4}{c|}{\bf AI techniques}&\multicolumn{6}{c|}{\bf Applications}\\
\cline{3-16}
\multicolumn{2}{|c|}{}&OMA & NOMA & RA & Switch & DL & RL &SL & FL& RIS & MIMO &THz&IoT&AdHoc&SAGIN \\
\shline
\multirow{4}{*}{} &\cite{dai2015game}  &\checkmark & & & & & \checkmark& & & & & &  &  &\checkmark\\
\cline{2-16}
\multirow{4}{*}{} & \cite{mertikopoulos2011distributed} &\checkmark & & & & & \checkmark &  & & &  & & \checkmark & &\\
\cline{2-16}
\multirow{4}{*}{} & \cite{mertikopoulos2016learning} &\checkmark & & & & &\checkmark &  & & & \checkmark & & & &\\
\cline{2-16}
\multirow{4}{*}{\bf{Centralized}} & \cite{singh2022machine} &\checkmark & & & &\checkmark & &  & & &  & & \checkmark & &\\
\cline{2-16}
\multirow{4}{*}{} & \cite{jiang2020multitask} & \checkmark& & & &\checkmark & &  & & &  & \checkmark&  & &\\
\cline{2-16}

\multirow{4}{*}{} & \cite{ye2020deepnoma} & & \checkmark& &  &\checkmark & &  & & &  & & \checkmark & &\\
\cline{2-16}

\multirow{4}{*}{} & \cite{huang2020deep} & &\checkmark & & &\checkmark & &  & & & \checkmark & &  & &\\
\cline{2-16}

\multirow{4}{*}{} & \cite{ahsan2021resource} & & \checkmark& & & &\checkmark  &  & & &  & &\checkmark  & &\\
\cline{2-16}

\multirow{4}{*}{} & \cite{zhong2021ai} & & \checkmark& & & \checkmark&\checkmark &  & &\checkmark &  & &  & &\\
\cline{2-16}

\multirow{4}{*}{} & \cite{ni2022star} & & \checkmark& & & & &  &\checkmark &\checkmark &  & &\checkmark & &\\

\cline{2-16}
\multirow{4}{*}{} & \cite{lin2023unsupervised} & & \checkmark& & &\checkmark & &  & &\checkmark &  &\checkmark & & &\\
\cline{2-16}
\multirow{4}{*}{} & \cite{huang2022deep} & & \checkmark& & &\checkmark & &  & &  &  & & & &\checkmark\\
\hline
\multirow{4}{*}{} & \cite{kumar2021adaptive}& & &\checkmark & & &\checkmark &  & & &  & & &\checkmark &\\
\cline{2-16}
\multirow{4}{*}{} & \cite{zhu2012achieving}& & &\checkmark & & &\checkmark &  & & &  & & &\checkmark &\\
\cline{2-16}
\multirow{4}{*}{\bf{Distributed}} & \cite{abyaneh2019intelligent}& & &\checkmark & & & & \checkmark & & &  & & & \checkmark&\\
\cline{2-16}
\multirow{4}{*}{} & \cite{edalat2019dynamically}& & &\checkmark & & & & \checkmark & & &  & & &\checkmark &\\
\cline{2-16}
\multirow{4}{*}{} & \cite{zhang2020enhancing}& & &\checkmark & & & \checkmark&  &\checkmark & &  & & & \checkmark&\\
\cline{2-16}
\multirow{4}{*}{} & \cite{wydmanski2021contention}& & &\checkmark & & &\checkmark &  & & &  & & &\checkmark &\\
\cline{2-16}
\multirow{4}{*}{} &\cite{ali2021federated} & & &\checkmark & & &\checkmark &  & \checkmark& &  & & &\checkmark &\\
\cline{2-16}
\multirow{4}{*}{} &\cite{yang2019machine} & & &\checkmark & & & & \checkmark  & & &  & &\checkmark & &\\
\hline
\multirow{4}{*}{} &\cite{cao2021ai}  &\checkmark & & \checkmark& \checkmark&\checkmark &\checkmark & & &\checkmark &  & &\checkmark  & &\checkmark\\
\cline{2-16}
\multirow{4}{*}{\bf{Hybrid}} & \cite{sharma2020hybrid}  &\checkmark & \checkmark& &\checkmark &\checkmark & & & & & \checkmark & & & &\\
\cline{2-16}
\multirow{4}{*}{} & \cite{qiao2018intelligent} &\checkmark & & \checkmark&\checkmark & & &\checkmark & & & \checkmark & & &\checkmark &\\
\cline{2-16}
\multirow{4}{*}{} & \cite{yang2018scalable} &\checkmark& &\checkmark&\checkmark &\checkmark & & & & &  & &\checkmark &\checkmark &\\
\cline{2-16}
\multirow{4}{*}{} &\cite{cao2019deep}  &\checkmark & &\checkmark &\checkmark & &\checkmark & & & &  & & \checkmark&\checkmark &\\
\cline{2-16}
\multirow{4}{*}{} & \cite{gomes2020automatic} &\checkmark & &\checkmark &\checkmark & &\checkmark& & & &  & &\checkmark &\checkmark &\\
\cline{2-16}
\multirow{4}{*}{} & \cite{chaieb2022deep} & \checkmark&\checkmark & & \checkmark& &\checkmark & & & & \checkmark & &\checkmark & &\\
\shline
\end{tabular}
\label{table_MAP}
\end{table*}

The AI-empowered hybrid MAC designs that alternately implement the centralized and distributed modes in a single frame by controlling the weight of each are detailed in Fig. \ref{figh}, which includes the following three cases. 

{\em Case 1}: As shown in Fig. \ref{figh}, the scheduled and contention transmissions are combined after the pilot transmissions and the computation, while allowing users to switch between them to meet different QoS requirements. Each frame is divided into four parts, namely the pilot period, the computation period, the scheduled transmission period, and the contention transmission period. According to the proposed design of the AI-empowered distributed MA schemes, the scheduled users transmit their data to the central controller in the scheduled transmission period. Then, based on the AI-empowered distributed MA scheme design, the unscheduled users (i.e., the unserved users who have sent their pilots and the new requesting users) transmit their data to the receivers during the contention transmission period. By dynamically switching between two transmission modes, this scheme is able to maintain the target rate. Note that the AI-empowered hybrid MA scheme switches to the AI-empowered centralized MAC scheme when the weight of w$_2$ is zero, or switches to the  AI-empowered distributed MA scheme when the weight of w$_1$ is zero.

{\em Case 2}: The contention requests and the scheduled transmissions are combined into a single frame. Each frame consists of the contention request period, the computation period, and the scheduled transmission period. The user sends a request to the central controller when a subchannel becomes available during the contention request period, and the central controller responds with an acknowledgment. Based on the received requests, the central controller sends the scheduling information to the users during the computation period. The scheduled users then transmit their data to the central controller within the scheduled transmission period.

{\em Case 3}: Similar to Case 2, the contention requests and the reserved data transmissions are combined into a single frame, and the computation is done at the user rather than at the central controller. When a subchannel becomes idle, the user occupies the channel, then calculates the required resources based on the RL model and sends a request to the central controller to reserve the channel resources for future data transmissions. The central controller sends feedback to the user once a request is registered. When the reserved transmission period arrives, the user transmits the data to the receiver in the reserved slots.

The comparison of the existing AI-empowered centralized, distributed, and hybrid MA protocols in terms of designs, AI approaches, and applications is summarized in Table \ref{tableC}.

\subsection{Discussions}\label{sec4d}
The designed AI-empowered MA protocols have been extensively studied in many works. However, the frame, protocol, and implementation designs combined with AI techniques still face some challenges.
\begin{enumerate}
\item The MA frame design allows nodes to access the channel using the same or a different MA scheme, which may lead to priority or fairness issues. Therefore, AI solutions for frame design (i.e., superframe or subframe) considering priority or fairness becomes a challenge. In addition, the changing networks and topologies can affect the frame design and lead to adaptive problems, and dynamic frame designs also introduce additional overhead and power consumption.

\item Different MA  protocols are designed to address different issues. For example, although the AI-empowered centralized MA protocol avoids collisions through scheduling, it does not fully address the complexity and overhead issues. On the other hand, the advanced AI algorithms are important for the collision avoidance of the distributed MA  protocol, especially for the high load requirements. Furthermore, the cooperation of different MA schemes brings new complex problems. For example, adaptive switching between the two for different network scenarios is required to take full account of the current and historical network states, and how to use AI techniques to select the optimal MAC scheme and reasonably set parameters for the selected MA  scheme are still open problems. In addition, the cross-layer AI-empowered MA protocol design involving the PHY layer and network layer optimization is a challenge for system performance improvement.

\item The MA implementation designs mainly focus on a single network and ignore the integration of different networks (e.g., SAGIN), so the flexible and compatible MA implementation design for different networks is still a challenge. For example, the characteristics of different networks, such as an unstable channel, rapidly changing channel states, low power, and long propagation delay need to be fully considered. Thus, the use of AI techniques to learn these characteristics to reduce collisions and improve MA performance remains to be solved. In addition, the structure-based designs require significant changes to the system framework, and then face challenges in hardware implementation and system compatibility.
\end{enumerate}

\begin{figure*}[!htp]
\centering
\includegraphics[width=5.5in]{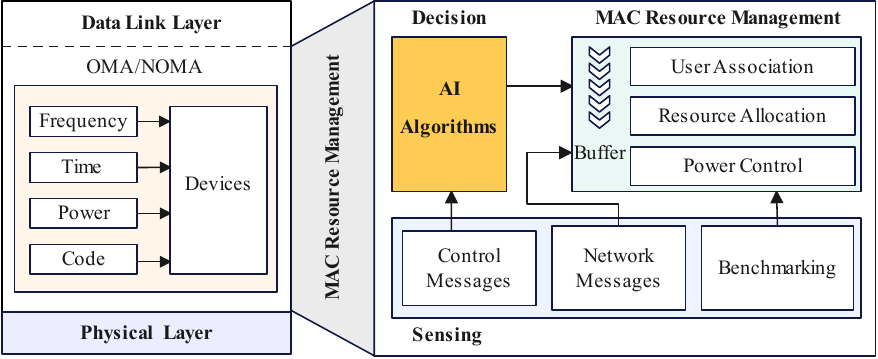}
\caption{Resource management for AI-empowered MA protocol optimization.}
\label{fig_d1}
\end{figure*}

\section{Optimization for AI-Empowered MA}\label{sec5}
Compared to typical optimization algorithms, AI-based approaches to MA protocol optimization are particularly efficient in reducing computational complexity and improving resource utilization, especially when dealing with complicated time-varying traffic or even unknown environments. Furthermore, AI technologies combined with MEC or WPT are considered to ensure timeliness and energy reduction for 6G systems. Therefore, the AI-enabled MA resource management is presented in Section \ref{sec5a}, the AI-enabled MA parameter adjustment is shown in Section \ref{sec5b}, the AI-enabled MA  access switching is depicted in Section \ref{sec5c}, and their discussions are given in Section \ref{sec5d}.

\subsection{Resource Management}\label{sec5a}

Resource management is an important aspect that significantly affects MAC performance and aims to achieve high spectrum and energy efficiency while suppressing inter/intra-cell interference. Resource management at the MAC layer is required to dynamically schedule users, allocate RBs and control transmit power levels. Traditional resource management can be performed using mathematical optimization theory, including convex optimization methods (e.g., interior-point method, Newton's method, Lagrangian duality method, etc.) and non-convex optimization methods (e.g., branch-and-bound (BnB), game theory, successive convex approximation, etc.). However, the mathematical optimization methods generally spend a lot of time and energy solving optimization problems in time-varying and unpredictable network environments, and the results obtained may be suboptimal. As shown in Fig. \ref{fig_d1}, AI algorithms are being explored in resource management at the MAC layer for user association, resource allocation, and power control, thereby reducing computational complexity and achieving approximate real-time solutions.

\subsubsection{User Association}
At the MAC layer, appropriate association schemes between users and the central controller (i.e., user scheduling by the central controller) are required for the designed MAC protocols. Considering the information related to SINR, load balancing, QoS requirements, and dynamic channel environment, some existing AI solutions are used to find the optimal user association scheme. For example, Pervez {\em et al.} \cite{pervez2017fuzzy} propose a distributed user cell association scheme. It uses a fuzzy Q-learning algorithm to learn the optimal bias values that guide users to associate with preferred cells. Li {\em et al.} \cite{li2017user} develop an online RL approach to solve the vehicle-BS association problem in vehicular networks. Initial RL and history-based RL are defined. In the initial RL phase, the BS decides the associated vehicles based on the reward defined to minimize the data rate deviation of the vehicles. In the history-based RL phase, the association schemes obtained in the initial RL phase enable the BSs to achieve load balancing as the environment changes dynamically. Challita {\em et al.} \cite{challita2018cellular} propose a DRL method to solve the UAV-cell association problem, thereby reducing the latency and interference of UAVs. Mauricio {\em et al.} \cite{mauricio2018low} use the K-means algorithm to divide mobile stations into spatially compatible clusters, thereby supporting multiple spatial streams per cluster. Raharya {\em et al.} \cite{raharya2020pursuit} propose a parallel pursuit learning algorithm to solve a joint optimization problem combining pilot assignment and multi-BS association. Unlike pursuit learning approaches that use only a binary reward, the proposed parallel pursuit learning algorithm allows a continuous reward to provide an accurate estimate of spectral efficiency. Furthermore, by introducing a heuristic solution in the pursuit-learning, the UE is associated with a BS with high network spectral efficiency while maintaining low complexity. Liu {\em et al.} \cite{liu2020user} develop an SL-based user association approach to support multi-connectivity in mmWave networks, where a graphical model is used to represent the mmWave user-BS association, and this association can be trained in a supervised manner with appropriate features obtained from both geographic location and topological information. To satisfy the required highly autonomous and decentralized nature, Dinh {\em et al.} \cite{dinh2021distributed} propose a distributed user-AP association method to maximize the long-term total rate under QoS constraints and AP load constraints, where the DRL algorithm is used to help APs make an association decision based on their local network state knowledge.

\subsubsection{Resource Allocation}
Spectral efficiency and energy efficiency are major concerns in 6G systems and therefore it is desirable to achieve highly efficient RB (i.e., time, frequency, code, and power) allocation for the designed MAC protocols to achieve these objectives. In this case, different AI techniques for RB allocation are presented to optimize the designed MAC protocols. In particular, the typical AI-based RB allocation for protocol optimization can be divided into four categories: DL-based resource allocation \cite{yang2019computation,yang2019computation,jiang2020multitask,cao2021reconfigurable1,cao2021reconfigurable2,lei2019learning, yang2019deep,zhang2020deep,huang2020deep,ali2021deep,camana2022deep,wu2023deep}, RL-based RB allocation \cite{liu2020situation,sah2022tdma,li2022dynamic,zhu2022dynamic,wang2022deep,he2019joint,shi2021deep,li2020resource,li2020learning,ahsan2021resource,ahsan2022reliable}, FL-based RB allocation \cite{zhong2022mobile,poposka2023resource,bouzinis2021wireless,zhong2022over}, and other SL/USL-based RB allocation \cite{gao2021machine,zhang2020energy}, which are detailed below.

DL techniques have been widely investigated in the protocol optimization of various MAC schemes due to their lower computational complexity and signalling overhead. For example, Yang {\em et al.} \cite{yang2019computation,yang2019computation} design a multi-task learning-based feedforward neural network model for the OMA/NOMA-based MEC system. The authors first use multi-task learning to address the joint optimization of offloading strategy and computational resource allocation, and also verify the effectiveness of the multi-task learning method in solving mixed-integer nonlinear programming (MINLP). To reduce computational complexity and mitigate multi-user interference, Jiang {\em et al.} \cite{jiang2020multitask} also use a multi-task learning model to achieve optimal multi-user hybrid beamforming in mm-wave massive MIMO OFDMA systems. Cao {\em et al.} \cite{cao2021reconfigurable1} propose a RIS-aided OMA protocol, and use a DNN-based multi-task learning model to predict the optimal transmission strategy of the proposed protocol in real-time, Cao {\em et al.} \cite{cao2021reconfigurable2} further investigate the RIS resource optimization of the RIS-aided MAC protocol relying on various AI techniques. In addition, most DL work has investigated resource optimization in NOMA systems. Lei {\em et al.} \cite{lei2019learning} adopt the DL model to tackle the NOMA resource allocation problems with high computational efficiency. Yang {\em et al.} \cite{yang2019deep} use a DNN-based method for NOMA systems to achieve real-time resource optimization. Zhang {\em et al.} \cite{zhang2020deep} use the DL model to deal with user association, subchannel, and power allocation in NOMA systems, thereby achieving higher energy efficiency and lower complexity. Huang {\em et al.} \cite{huang2020deep} propose a DNN algorithm for MIMO-NOMA networks to address the dynamic power allocation problem. Ali {\em et al.} \cite{ali2021deep} develop a DNN method for joint uplink-downlink NOMA optimization, where the DNN model is used to solve joint power loading problems at source and relay nodes, thereby maximizing the sum rate. As a further step, Camana {\em et al.} \cite{camana2022deep} apply the DNN model in RSMA to provide the optimal precoder vector with lower computational complexity. To address dynamic mobility issues, Gao {\em et al.} \cite{gao2022machine} propose a DL-based algorithm for beam management and optimization in mmWave NOMA communications. To achieve accurate high-dimensional CSI and reduce signalling overhead in RIS-aided THz massive MIMO systems, Wu {\em et al.} \cite{wu2023deep} propose a transformer-based data-driven RIS reflecting network to realize passive precoding at the RIS, and further design a model-driven deep unfolding active precoding network for RSMA digital active precoding at the BS.

To address resource allocation problems in a dynamic or uncertain environment, RL models are used to allocate RB efficiently from their historical experience. Considering a time-division-duplex (TDD)-based multi-dimensional intelligent multiple access, Liu {\em et al.} \cite{liu2020situation} use a deterministic policy gradient algorithm, a model-free DRL technique, to optimize the multi-dimensional radio resource allocation in real-time with faster convergence speed. Sah {\em et al.} \cite{sah2022tdma} develop an RL algorithm to provide an optimal scheduling policy in TDMA systems, and use Q-learning to find the optimal cost to improve throughput and reduce latency. To improve spectral utilization, Li {\em et al.} \cite{li2022dynamic} use a DRL algorithm to learn the best access strategy in FDMA and NOMA systems. Zhu {\em et al.} \cite{zhu2022dynamic} propose a DRL algorithm to support efficient bandwidth allocation in OFDMA passive optical networks, where slots, subcarriers, and modulation formats are dynamically allocated using the DRL model. Wang {\em et al.} \cite{wang2022deep} design a multi-agent deep Q-network algorithm for beamforming coordination and resource allocation in multi-cell MISO-OFDMA systems, where the proposed algorithm enables the new agents to predict the resource allocation based on their experience and the learning results of the pre-trained agents. Besides the above conventional OMA schemes, extensive RL approaches have been studied for resource allocation in NOMA systems. To fully exploit the advantages of NOMA techniques, He {\em et al.} \cite{he2019joint} propose an attention-based neural network (ANN)-based DRL algorithm to optimally allocate power and channel resources in the multi-carrier NOMA system. Considering the continuity of resource allocation, Shi {\em et al.} \cite{shi2021deep} propose a deep deterministic policy gradient (DDPG) algorithm for multi-dimensional resource allocation in NOMA communication. To reduce energy consumption and computation delay, Li {\em et al.} \cite{li2020resource} use the mean-field DDPG algorithm to achieve power and computation resource allocation in NOMA-based MEC systems. Li {\em et al.} \cite{li2020learning} also propose a DQN-based constrained Markov decision process to optimize the resource allocation for PDMA scheme. To increase the throughput of NOMA-IoT networks, Ahsan {\em et al.} \cite{ahsan2021resource} develop DRL and SARSA learning algorithms to explore the fair resource allocation in NOMA while balancing RB and network traffic. Ahsan {\em et al.} \cite{ahsan2022reliable} further consider reliable resource sharing in NOMA-URLLC, and propose a DRL approach to provide long-term resource allocation based on instantaneous reward feedback.

To improve the efficiency and effectiveness of RL in RIS-aided NOMA systems, Zhong {\em et al.} \cite{zhong2022mobile} adopt a FL-aided DRL algorithm to optimize RIS design and power allocation policy, where FL is used to assist multiple agents to exchange experience, and these cooperated agents can obtain a reward gain compared to the independent agents. To ensure fair resource sharing among wireless stations, Poposka {\em et al.} \cite{poposka2023resource} use an FL model to transmit local model parameters to the base station and achieve fair resource allocation with NOMA. Here, only a single wireless station is allowed to transmit at maximum power, while the other wireless stations transmit at lower power. Bouzinis {\em et al.} \cite{bouzinis2021wireless} introduce and optimize a compute-then-transmit NOMA for wireless federated learning (WFL) networks, where the users simultaneously transmit the locally trained parameters to the server, and the server jointly optimizes computation and communication resources to reduce the overall delay during a WFL communication round. Zhong {\em et al.} \cite{zhong2022over} study over-the-air federated multi-task learning (OA-FMTL) for the MIMO-MAC optimization to reduce the impact of inter-task interference on the FL performance. To maximize the sum rate, Gao {\em et al.} \cite{gao2021machine} propose a K-means algorithm to optimize the RIS phase shift matrix and power allocation coefficient vector in RIS-aided MISO NOMA systems. To improve the energy efficiency in THz-NOMA-MIMO systems, Zhang {\em et al.} \cite{zhang2020energy} use an enhanced K-means ML algorithm for solving user clustering and power allocation problems.

\subsubsection{Power Control}
Optimal power control and management are beneficial to reduce interference and increase system throughput when optimizing the AI-empowered MA protocol. Existing AI methods for power control and management in different MA protocols are detailed below.

To reduce primary user interference and avoid signalling overhead in CNN, Galindo {\em et al.} \cite{galindo2010distributed} introduce a decentralized Q-learning algorithm to learn an optimal power control policy and manage the aggregated interference. Bennis {\em et al.} \cite{bennis2013self} formulate the power control and carrier selection as a non-cooperative game among femto BSs, and propose an RL algorithm to help FBSs to reach a Logit equilibrium. To mitigate inter-cell interference and maximize the system performance, Simsek {\em et al.} \cite{simsek2014learning} propose a two-stage RL formulation for learning the optimal cell range expansion bias, power control schemes, and frequency band allocation. Asheralieva {\em et al.} \cite{asheralieva2016autonomous} develop a fully autonomous multi-agent Q-learning to obtain the best channel and power control strategies in heterogeneous cellular networks, achieving fast convergence and near-optimal performance. Xiao {\em et al.} \cite{xiao2017reinforcement} propose a Q-learning-based power allocation algorithm for the downlink NOMA transmission without knowing the jamming and radio channel parameters, and a Dyna architecture based on real anti-jamming transmission experience is used to accelerate the convergence of the Q-learning-based power allocation and improve NOMA efficiency. Chu {\em et al.} \cite{chu2019power} apply an actor-critic DQN-based RL algorithm in an EH-MA system for orthogonal channel access and continuous power control, while an LSTM-based algorithm is used for battery state learning. Bouzinis {\em et al.} \cite{bouzinis2021wireless} propose an online stochastic learning algorithm for delay-optimal power control and sub-band allocation in the OFDMA uplink system. To deal with Q-learning model overestimation issues, Fayaz {\em et al.} \cite{fayaz2021transmit} propose a multi-agent DQN-based grant-free NOMA algorithm to optimize transmit power levels for open-loop power control. To ensure UAV flight safety and satisfy the sum rate requirement, Zhao {\em et al.} \cite{zhao2022ris} propose a sample-efficient DRL algorithm for power control in RIS-NOMA-aided UAV systems. To balance system performance and computational complexity, Zhang {\em et al.} \cite{zhang2023unsupervised} propose a USL algorithm for power control strategies in cell-free massive MIMO systems. It can be learned from \cite{galindo2010distributed,bennis2013self,simsek2014learning,asheralieva2016autonomous,xiao2017reinforcement} that distributed Q--earning-based power control strategies perform better in channel access in CRNs and heterogeneous networks. According to  \cite{chu2019power,bouzinis2021wireless}, the performance of OMA can be improved by applying RL algorithms in power control. In addition, using DRL algorithms to achieve real-time power allocation in NOMA systems is also a potential way to reduce the computational complexity and improve NOMA performance \cite{fayaz2021transmit,zhao2022ris,zhang2023unsupervised}.

\begin{figure*}[!htp]
\centering
\includegraphics[width=5.5in]{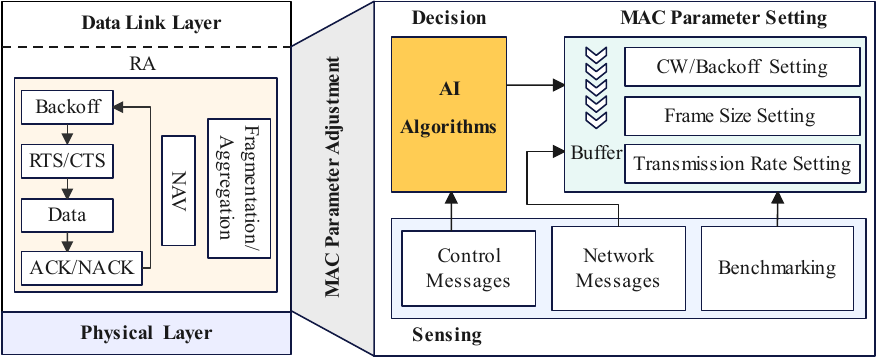}
\caption{Parameter adjustment for AI-empowered MA protocol optimization.}
\label{fig_d2}
\end{figure*}

\subsection{Parameter Adjustment}\label{sec5b}
Employing AI solutions for IEEE 802.11, MA protocols can be optimized by dynamically adjusting their access parameters, including CW/backoff, frame size, and transmission rate setting, as shown in Fig. \ref{fig_d2}, to reduce MA collisions and improve MA performance.

\subsubsection{CW and Back off Setting}

According to \cite{ ali2018deep, coronado2020improvements,kumar2021adaptive, abyaneh2019intelligent, edalat2019dynamically, wydmanski2021contention,zhu2012achieving, zhang2020enhancing, amuru2015send, ali2020performance}, applying RL-based and SL-based methods, the CW/backoff values of MA protocols can be optimized to improve throughput by reducing both collisions and idle periods. For example, the existing RL \cite{ali2018deep, coronado2020improvements, kumar2021adaptive, wydmanski2021contention,zhu2012achieving, zhang2020enhancing, amuru2015send, ali2020performance} and SL \cite{abyaneh2019intelligent,edalat2019dynamically} solutions for IEEE 802.11 standards mainly focus on improving throughput by adjusting the CW/backoff value of the basic 802.11 MAC protocol, where a large CW or a small CW value may reduce throughput due to the increasing idle slots or increasing collision probability, and how to utilize different AI techniques to set an optimal CW value has been widely studied. For example, Zhu {\em et al.} \cite{zhu2012achieving} implement an adaptation-based programming algorithm to halve or leave unchanged the CW size depending on the transmission state. Amuru {\em et al.} \cite{amuru2015send} propose a post-decision state-based learning to use the previous knowledge of the CW setting to increase the convergence speed. Based on the channel observation-based scaled backoff protocol, Ali {\em et al.} \cite{ali2018deep} optimize the CW value according to the learned channel collision probabilities. Edalat {\em et al.} \cite{edalat2019dynamically} achieve a balance between small CW and large CW using fixed-share SL. Abyaneh {\em et al.} \cite{abyaneh2019intelligent} adopt SL to balance the minimum CW size to achieve fairness. To ensure access priority, Ali {\em et al.} \cite{ali2020performance} implement QL to infer the network density and set the CW value. Zhang {\em et al.} \cite{zhang2020enhancing} combine FL and DRL models to set the CW value to achieve fairness. Kumar {\em et al.} \cite {kumar2021adaptive} propose DQL to train minimum CW selection to achieve per-user fairness. Wydmanski {\em et al.} \cite{wydmanski2021contention} use DRL to learn the optimal CW value setting to maintain a stable throughput as the number of users increases. For distributed channel access (EDCA) in the 802.11e amendment, arbitration inter-frame space (AIFS) and transmission opportunity (TXOP) are included as the extended MAC parameters, and the joint optimization of AIFS and CW is discussed to balance delay and throughput. Coronado {\em et al.} \cite{coronado2020improvements} select the optimized combination of CW and AIFS via RL. The AIFS and CW values are first set relying on decision tree algorithms, and then the optimal combination for AIFS and CW is predicted. In EDCA, the optimal CW setting is adaptive to the priority of the traffic.

\subsubsection{Frame Size Setting}
The frame size setting has a direct impact on the efficiency of MA protocols in terms of useful data transmitted and overhead. In general, SL-based and RL-based approaches have been used to dynamically optimize the frame size for IEEE 802.11 networks to improve throughput. For example, some works have developed slot selection methods using RL models. In particular, Lee {\em et al.} \cite{lee2020collision} select available time slots using RL. Kihira {\em et al.} \cite{kihira2020adversarial} perform the best time slot-based transmission policies via multi-agent cooperative RL, and Guo {\em et al.} \cite{guo2022multi} propose a multi-agent RL algorithm to select time slots for the optimal channel access behaviour. Several learning mechanisms are presented to dynamically select the frame size, including SL \cite{lin2009machine, coronado2020adaptive, coronado2020aios, khastoo2020neura} and RL \cite{karmakar2019online}. Lin {\em et al.} \cite{lin2009machine} jointly select both the frame size and the CW value to address channel collisions, where an ANN model is trained with frame size-throughput patterns to indicate the optimal frame size and the CW value. Karmakar {\em et al.} \cite{karmakar2019online} design the frame size selection using the RL model to achieve the energy-throughput trade-off. Coronado {\em et al.} \cite{coronado2020adaptive} implement a random forest regressor (RFR) model to configure the frame size and the modulation and coding scheme settings. Coronado {\em et al.} \cite{coronado2020aios} further develop an RFR-based frame size selection method for software-defined WLANs to maximize the goodput of each user. This solution defines a set of suitable aggregation levels based on an online learning algorithm and selects the optimal level from this set by using fuzzy logic to estimate which frame size would have the lowest FER. Hassani {\em et al.} \cite{hassani2021quick} use a logistic regression estimator model to determine the optimal level of aggregation with low computational complexity. Khastoo {\em et al.} \cite{khastoo2020neura} predict the optimal modulation and coding scheme setting using the ANN model, where the optimal frame aggregation level is estimated using a previously trained model. Chen {\em et al.} \cite{chen2021experience} select the modulation and coding scheme (MCS) values to achieve the best goodput using the double deep Q-network model.

\subsubsection{Transmission Rate Setting}
Configuring the link by selecting the MAC parameters becomes a key factor in achieving the optimum throughput for the given network and CSI. In particular, rate adaptation for the selection of MCS values for each transmission plays an important role in the link configuration. Considering user mobility or interference in dynamic access scenarios, rate adaptation is required to address the trade-off between transmission errors and channel utilization using AI models, especially considering varying CSI. The optimal transmission rate predictions are obtained by relying on SNR, or ACK/NACK feedback, where SNR solutions are suitable for real-time mobility services, and ACK and NACK feedback solutions are more accurate. To improve throughput, SNR-based or ACK/NACK-based predictions are usually achieved by applying AI techniques. There is a wide range of work that is intrinsically linked to parameter configuration using DL-based and RL-based methods. For example, Joshi {\em et al.} \cite{joshi2008sara} propose a stochastic learning automata-based method for selecting available data rates according to the reward function with respect to throughput. To improve the aggregate throughput, Wang {\em et al.} \cite{wang2013dynamic} propose a legacy auto-rate fallback algorithm to predict the data rate based on the thresholds; these thresholds are adjusted by an ANN model and affect the number of ACKs received. Kurniawan {\em et al.} \cite{kurniawan2018machine} achieve transmission rate selection based on the channel condition, where the trained SL model for the optimal MCS level selection relies on the selected characteristics of the preamble of an 802.11 frame. Cho {\em et al.} \cite{cho2021reinforcement} use a Q-learning model for the MCS level selection according to the number of ACKs received. Karmakar {\em et al.} \cite{karmakar2020deep} use the DNN model to adjust channel bonding, transmission rate, and frame size to improve throughput. Moon {\em et al.} \cite{moon2021neuro} enable ML agents to self-configure link parameters based on gathered experience. Chen {\em et al.} \cite{chen2021experience} apply the DDQN model to select the transmission rate and MCS levels considering the available bandwidth and spatial streams, where goodput is the reward, and includes prioritized training, history-based initialization, and adaptive training interval. Ali {\em et al.} \cite{ali2021federated} propose a federated RL-based model for link parameter configuration with faster learning convergence.

\begin{figure*}[!htp]
\centering
\includegraphics[width=5.5in]{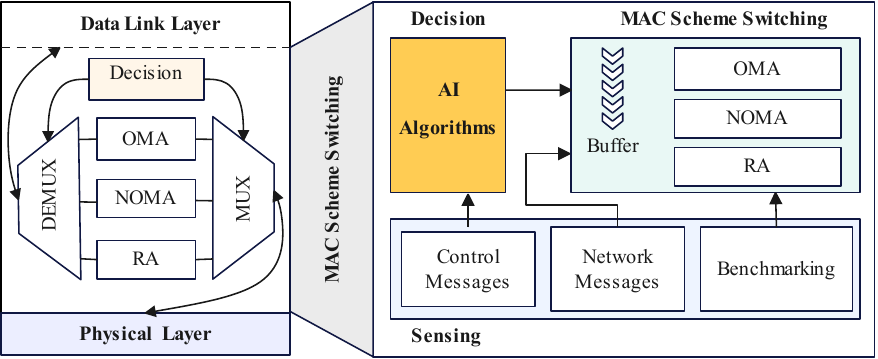}
\caption{Scheme switching for AI-empowered MA protocol optimization.}
\label{fig_d3}
\end{figure*}

\subsection{Access Switching}\label{sec5c}
AI solutions for optimal switching between OMA, NOMA, and RA are shown in Fig. \ref{fig_d2}, where DL or RL models are used to select an optimal MA scheme for different uses in dynamic scenarios, thereby achieving access switching between different MA schemes for different QoS requirements. Such as switching between TDMA, FDMA and CDMA schemes \cite{salajegheh2007hymac,shu2009energy}, switching between OMA and NOMA schemes \cite{sharma2020hybrid,zhao2022drl,chaieb2022deep}, switching between OMA and RA schemes \cite{rhee2005z,qiao2018intelligent,yang2018scalable,cao2019deep,cao2021reconfigurable2}, and switching between NOMA and RA schemes.

Traditionally, the optimization of hybrid MAC scheme selection is based on numerical algorithms. For example, Rhee {\em et al.} \cite{rhee2005z} propose a Z-MAC that switches between TDMA and CSMA while compensating for their weaknesses. Similarly, Zhang {\em et al.} \cite{zhang2010hybrid} investigate the optimal coexistence of reservation-based and contention-based MA schemes and verify the trade-off between them through numerical analysis. Salajegheh {\em et al.} \cite{salajegheh2007hymac} achieve an optimal switching between TDMA and FDMA schemes for energy-efficient collision-free channel access in WSNs. Shu {\em et al.} \cite{shu2009energy} use a heuristic algorithm to achieve an optimal switching between TDMA and CDMA to improve energy efficiency in WSNs. Pedersen {\em et al.} \cite{pedersen2016flexible} optimize the fundamental tradeoffs between spectral efficiency, latency, and reliability by implementing flexible MA schemes. However, these existing numerical methods for MA switching result in high computational complexity and low spectral and energy efficiency.

In order to improve spectral efficiency and energy efficiency while significantly reducing computational complexity, AI-based methods are used to achieve optimal switching between different MA schemes. In particular, Qiao {\em et al.} \cite{qiao2018intelligent} use ML methods to select MA schemes according to the inherent behaviour and external environment. Then the optimal MA scheme is learned between competitive-based and non-competitive-based protocols to adapt to the network condition. Yang {\em et al.} \cite{yang2018scalable} implement NNP-based ML algorithms for a scalable MAC framework to achieve an optimal coexistence between schedule-based and contention-based MA schemes, by adjusting the length of the contention access period.  It also achieves the trade-off between scheduled-based and contention-based MA schemes. To adapt to the dynamic network environment and user characteristics, Cao {\em et al.} \cite{cao2019deep,cao2021reconfigurable2} use the DL and DRL approaches to achieve the trade-off between reserved-based and contention-based MA schemes, Gomes{\em et al.} \cite{gomes2020automatic} also employ an RL approach to achieve the switching between reserved-based and contention-based MA schemes according to the current network demand. To support a highly overloaded multi-user system, Sharma {\em et al.} \cite{sharma2020hybrid} propose two DNN-based DL models for selecting between OFDMA and SCMA schemes, where one DNN-based DL model is used for symbol detection of near users adopting OFDMA, and the other one for symbol detection of far users adopting SCMA. Simulation results demonstrate that the DNN-based detection outperforms the message exchange-based algorithm and the SCI-based detection. To overcome the computational complexity of model-based approaches, Chai {\em et al.} \cite{chaieb2022deep} propose several parallel DNN algorithms to provide resource allocation solutions, including user association, power allocation, sub-channel assignment, and MA selection. Zhao {\em et al.} \cite{zhao2022drl} propose a DDPG-based DRL algorithm for resource allocation and IIoT device orchestration policy in NOMA communications.

\begin{figure*}[!htp]
\centering
\includegraphics[width=7in]{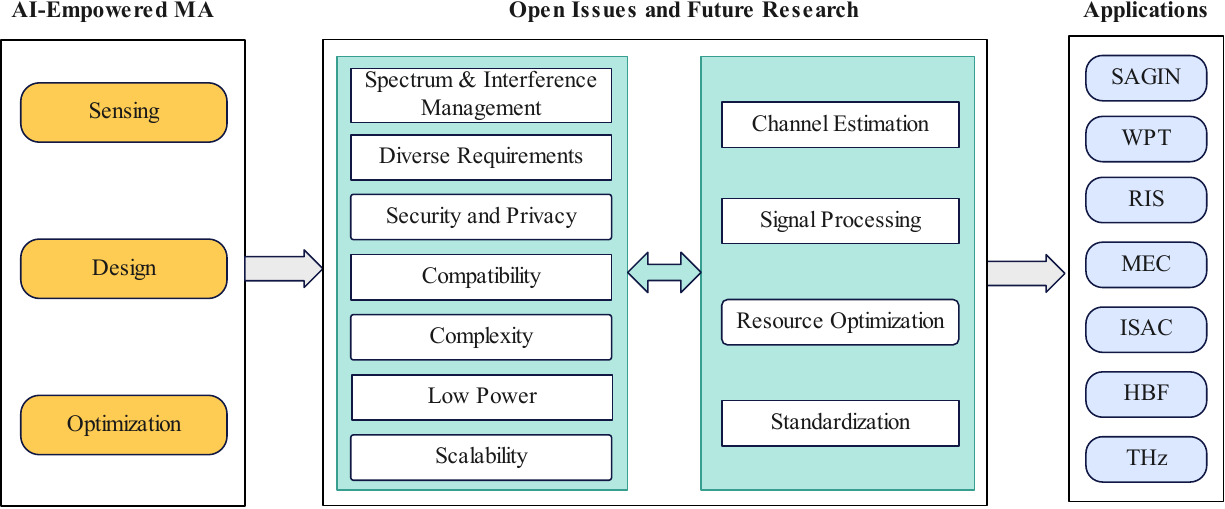}
\caption{Open issues and future researches for AI-empowered MA.}
\label{fig_a}
\end{figure*}

\subsection{Discussions}\label{sec5d}

From the multitude of papers dealing with the optimization of designed AI-empowered MA protocols, we identify the following discussions related to the different AI techniques for MA functions.

\begin{enumerate}
\item In 6G systems, traditional mathematical methods for solving MA protocol optimization are limited due to the large, diverse, dynamic, and heterogeneous nature of wireless networks. In addition, traditional mathematical optimization requires complete or quasi-complete knowledge of the wireless environment, which can lead to higher overhead. In contrast to traditional mathematical methods, AI-based solutions for MA protocol optimization are particularly efficient when dealing with complicated time-varying, or even unknown environments. However, AI-based solutions may lose their advantages when faced with static or quasi-static scenarios, as the theoretical optimization cannot be strictly guaranteed and the low computational complexity may be offset by the training required. Therefore, advanced AI-based solutions, such as unsupervised continuous learning algorithms are needed for complicated dynamic scenarios, including both mobility and static scenarios, to provide accurate prediction with less knowledge of the wireless environment.

\item Most of the existing AI solutions mainly focus on the specific optimization of MA protocol parameters under certain conditions, but lack an overall perspective of the network functioning, especially without considering the optimization criteria for cross-layers simultaneously. Therefore, AI solutions in this area face significant challenges in dealing with the large number of parameters involved in the MA protocols.

\item The performance, applicability, and benefits of AI solutions for MA protocol optimization have not been thoroughly investigated. In addition, the impact of realistic network status parameters on MA protocol optimization while exploiting the characteristics of the growing number of intelligent users needs to be further investigated. Furthermore, it is seen that user grouping is a potential direction for MA protocol design, which has further implications for resource optimization. However, exploiting the specific characteristics of users and classifying them into different groups for MA protocol design may lead to complex user association problems, making AI-based optimization for MA protocols inefficient, especially when the number of users and groups increases rapidly.
\end{enumerate}

\section{Open Issues and Outlook}\label{sec6}
The revenue of AI solutions in the 6G system is expected to facilitate many activities to develop MA protocol designs and optimizations that control various aspects of channel access. Although a lot of research has been done on AI-empowered MA protocols, some issues are still open, and further research efforts in this area are summarized in Fig. \ref{fig_a}. In this section, open issues and future research areas are summarized in Sections \ref{sec6a} and \ref{sec6b}, respectively.

\subsection{Open Issues}\label{sec6a}
\subsubsection{Spectrum and Interference Management}
Although comprehensive AI-empowered MA protocol designs and optimizations can improve spectral efficiency with low computational complexity and low overhead, it is still essential to address the limited 6G spectrum resources and interference issues, and to develop innovative spectrum management techniques. In 6G systems, facing high frequency bands (such as mmWave, THz, etc.), highly dynamic environments, complex heterogeneous networks, and exploding smart terminals, it is necessary to investigate 1) how to use advanced AI solutions for spectrum sensing, spectrum sharing, and SIM to reduce computational complexity, 2) how to allocate abundant spectrum resources to maximize spectral resource utilization, and 3) how to use interference cancellation methods to improve spectral efficiency.

\subsubsection{Diverse Requirements}
Different applications in the 6G system must take into account the different requirements of different users. For example, different traffic with periodic, bursty, elastic, random, or peak-to-average rates, different states with mobile, static, or quasi-static, different power consumption with active or passive, and different levels of intelligence. The AI-empowered MA protocol designs and optimizations are required to meet these diverse requirements while increasing access efficiency and supporting prioritized access for mission-critical applications. In addition, given the diverse characteristics of the scenarios, AI solutions are not superior because their optimizations may not be fully stringent, and AI solutions for access in static or quasi-static environments may offset their advantages in computational complexity due to the training required.

\subsubsection{Security and Privacy}
Security and privacy issues for AI-empowered MA protocol designs and optimizations are of great importance in 6G systems due to their human-centric characteristics. Unlike the traditional security challenges in 5G, the related security issues in 6G are not only related to decentralization, transparency, data interoperability, and privacy vulnerabilities at the PHY layer but also include access security issues (e.g., AI security/privacy, interference, intrusion, etc.) at the MAC layer when big data and AI technologies are widely used to address spectrum sensing, protocol design and optimization for MA systems.

\subsubsection{Compatibility}
In general, MA protocols are designed for networks following a fixed standard and for specific transceiver hardware. As technologies evolve from 5G to 6G, especially with the introduction of AI methods in 6G systems for wireless network management and applications, the existing entities in 6G systems will be able to cope with significant changes in the hardware settings, which may require a more complex architecture (e.g., terminals or servers are equipped with AI models for signal processing, computation, and localization). In this case, AI-empowered MA protocol designs and optimizations will be developed to ensure compatibility with these changes. In addition, the 6G system, with its large, diverse, and heterogeneous characteristics, makes compatibility to be more complex.

\subsubsection{Complexity}
With the development of LEO satellites and UAV technologies, the emerging SAGIN has a large vertical extension. The interaction and cooperation between different networks must lead to more complex design and optimization problems, which are extremely difficult to support sophisticated network functionalities, thereby posing a great challenge to AI-empowered MA protocol. In addition, the emergence of RIS technologies further increases the complexity of AI-empowered MA protocol design and optimization, as it involves accessing and designing RISs, especially when multiple RISs are involved. Finally, the different AI approaches used in MA protocol design to achieve specific benefits lead to complexity in deployment and implementation. Therefore, striking a balance between performance and complexity in AI-empowered MA protocol design and optimization is an important issue.

\subsubsection{Lower Power}
Low power consumption is an important issue in the design and optimization of AI-empowered MA protocols, especially for low-power terminals. However, most existing studies focus on MA performance and computational complexity from the perspective of spectrum utilization, without paying much attention to power consumption. Power issues are particularly challenging for MA protocol design and optimization as terminals become more intelligent. For example, network entities may be required to identify, invoke, or maintain AI functionalities, which may lead to additional power consumption due to the use of AI models. Therefore, AI-empowered MA protocols need to be developed with a focus on low-power, long-term spectrum sensing, accurate CSI estimation, low-complexity design and optimization, and lightweight AI models. This remains an open issue.

\subsubsection{Scalability}
AI-empowered MA protocol designs and optimizations should be able to adapt to the dynamic changes of networks. For example, it is crucial to research how to design MA protocols that can accommodate the different characteristics of dynamic networks and enable them to switch between different networks. It is also important to investigate how MA protocol designs and optimizations can adapt to the different types of intelligent networks. Additionally, these designs need to adapt to new physical layer (PHY) techniques. It is crucial to understand how to enable network participants to effectively comprehend the operational mechanism of signal processing techniques.

\subsection{Future Research Areas}\label{sec6b}

The current research on AI-empowered MA protocol design and optimization in 6G systems is still in its early stages, showing great potential in various application areas such as SAGIN, WPT, RIS, MEC, ISAC, HBF, and THz. For instance, combining AI with WPT and/or RIS can help solve power consumption issues, while combining with MEC and/or ISAC can reduce complexity. Additionally, combining with HBF, THz, and/or ISAC can address spectrum and interference management challenges, and combining with WPT, SAGIN, RIS, and/or ISAC can tackle diverse requirements, compatibility, and adaptability issues. Moreover, combining with HBF and/or RIS can enhance security and privacy.

Furthermore, recent AI innovations like generative AI and large language models (LLM) are bringing about a paradigm shift and have been applied to wireless communications. For example, generative AI can solve optimization and security issues by learning complex data distributions, while LLM can manage complex networking tasks by continuously acquiring updated network knowledge through an application programming interface. The success of these techniques lies in their scalability, accuracy, and generalization ability. Therefore, the ongoing development of advanced AI techniques offers a significant opportunity to address MA problems in the future. Table \ref{table} summarizes studies from various research areas that have identified potential technologies to address these open issues.

\begin{table*}[!htp]
\newcommand{\tabincell}[2]{\begin{tabular}{@{}#1@{}}#2\end{tabular}}
		\small
		\centering
			\renewcommand{\arraystretch}{1.2}
			\captionsetup{font={small}} 
			\caption{\scshape Potential research areas for AI-empowered MA} 
{
			\label{table}
			%\footnotesize
			\centering  
			\begin{tabular}{|m{0.2\textwidth}|m{0.75\textwidth}|}  
				\shline
				%\rowcolor[gray]{0.9}
			   \textbf{Future research areas} &\textbf{Directions}\\
				\shline  
			   \multirow{4}{*}{{Resource Optimization}} &  $\star$ AI-empowered MA protocol optimization with THz, RIS, HBF, MEC, WPT, etc. \\ 
			   \multirow{4}{*}{} &   $\star$ Advanced AI-assisted optimization algorithms for MA protocol \\ 
                        \multirow{4}{*}{} &   $\star$ The integration of sensing, communication, and computing for AI-empowered MA optimization\\ 
                        \multirow{4}{*}{} &   $\star$ Distributed optimization algorithms for AI-empowered MA in ubiquitous connectivity \\ 
	                 \hline  
			   \multirow{3}{*}{{Interference Management}} &  $\star$ SIC techniques of enhanced NOMA in 6G\\ 
			   \multirow{3}{*}{} &   $\star$ Co-channel interference and loss reduction by advanced beamforming or software-defined HBF \\ 
                   \multirow{3}{*}{} &   $\star$ Accurate spectrum sensing technologies for MA protocol\\ 
                        \hline  
			   \multirow{2}{*}{{Channel Estimation}} &  $\star$ CSI analysis in the complexity wireless environment or imperfect channel \\ 
			   \multirow{2}{*}{} &   $\star$ Adaptive AI-empowered MA designs considering dynamic CSI \\ 

                        \hline  
			   \multirow{2}{*}{{Signal Processing}} &  $\star$ AI-assisted detection, coding, and modulation algorithms for MA \\ 
			   \multirow{2}{*}{} &   $\star$ Advanced signal processing technologies for MA interference mitigation and complexity reduction  \\ 
                        \hline  
			   \multirow{3}{*}{{Complexity}} &  $\star$ Efficient MA protocol design without complexity hardware cost and overhead \\ 
			   \multirow{3}{*}{} &   $\star$ Computational complexity of MA protocol optimization\\ 
                   \multirow{3}{*}{} &   $\star$ Hybrid MA design in the complexity SAGIN systems. \\ 
                        \hline  
			   \multirow{3}{*}{{Security}} &  $\star$  The security and privacy of MA designs  \\ 
			   \multirow{3}{*}{} &   $\star$ The security and privacy of AI algorithms enabled in MA \\ 
                   \multirow{3}{*}{} &   $\star$ Data sets security in AI-empowered MA \\ 
                   \hline  
			   \multirow{2}{*}{{Standardization}} &  $\star$ The interplay design between MA protocol and 6G new technologies \\ 
			   \multirow{2}{*}{} &   $\star$ The compatibility, scalability, sustainability of AI-empowered MA designs for 6G and beyond \\ 
			   \shline
			\end{tabular} 
} 
	\end{table*}

\section{Conclusion}\label{sec7}
In this paper, we have provided one of the first comprehensive surveys on AI methods for designing and optimizing MA protocols in 6G systems. In particular, we have reviewed the various AI-empowered MA protocols that improve spectral and energy efficiency while meeting the new features and requirements of 6G networks. We then identified a unified framework model that integrates the AI-enabled sensing, design, and optimization models into MA protocols. Next, we introduced AI-enabled spectrum sensing, consisting of spectrum sharing and interference management, to improve MA performance and reduce collisions. We also have investigated existing AI-empowered MA protocols in terms of flexible frame structure, scalable access scheme, and dynamic implementation designs. To further improve MA efficiency, we have explored the optimization of AI-empowered MA protocols by enabling AI algorithms to perform resource management, parameter adjustment, and access switching at the MAC layer. In addition, we have discussed some critical challenges for spectrum sensing, protocol design, and performance optimization based on AI techniques, respectively. Last, we have presented some open issues and potential applications of AI-empowered MA protocols in 6G systems. The future of AI-empowered MA protocols will inevitably rely on the perfect interplay between AI technologies and MAC layer mechanisms, and so this paper provides a stepping stone toward understanding the protocol design and optimization needed to develop such highly efficient AI-empowered MA.

%\section*{Acknowledgments}
%This should be a simple paragraph before the References to thank those individuals and institutions who have supported your work on this article.

\bibliographystyle{IEEEtran}
\bibliography{REF}

%\newpage

\vspace{10pt}

\vfill

\end{document}